\documentclass[12pt]{article}
\usepackage{epsfig,amsfonts,amsbsy}
\newtheorem{theorem}{Theorem}[section]
\newtheorem{lemma}[theorem]{Lemma}

\newtheorem{pro}[theorem]{Proposition}

\makeatletter
\@addtoreset{equation}{section}
\makeatother

\newcommand{\proof}{\par\bigskip\noindent{\em Proof:\ }}

\newcommand{\qed}{~\rule{2mm}{2mm}\par\bigskip}
\newcommand{\ret}{\nonumber\\}

\newcommand{\norm}[1]{\left\Vert#1\right\Vert}
\newcommand{\rbk}[1]{\left(#1\right)}
\newcommand{\sqbk}[1]{\left[#1\right]}
\newcommand{\cbk}[1]{\left\{#1\right\}}

\newcommand{\set}[2]{\left\{#1\,\Bigl|\,#2\right\}}
\newcommand{\sumtwo}[2]%
{\mathop{\sum_{#1}}_{#2}}
\newcommand{\sumthree}[3]%
{\mathop{\mathop{\sum_{#1}}_{#2}}_{#3}}
\newcommand{\sumfour}[4]%
{\mathop{\mathop{\mathop{\sum_{#1}}_{#2}}_{#3}}_{#4}} 
\newcommand{\suptwo}[2]%
{\mathop{\sup_{#1}}_{#2}}
\newcommand{\supthree}[3]%
{\mathop{\mathop{\sup_{#1}}_{#2}}_{#3}}
\newcommand{\supfour}[4]%
{\mathop{\mathop{\mathop{\sup_{#1}}_{#2}}_{#3}}_{#4}} 
\newcommand{\inftwo}[2]%
{\mathop{\inf_{#1}}_{#2}}
\newcommand{\infthree}[3]%
{\mathop{\mathop{\inf_{#1}}_{#2}}_{#3}}
\newcommand{\inffour}[4]%
{\mathop{\mathop{\mathop{\inf_{#1}}_{#2}}_{#3}}_{#4}} 

\newcommand{\calB}{{\cal B}}

\newcommand{\calE}{{\cal E}}

\newcommand{\calI}{{\cal I}}

\newcommand{\calK}{{\cal K}}

\newcommand{\La}{\Lambda}
\newcommand{\up}{\uparrow}
\newcommand{\dn}{\downarrow}

\newcommand{\gtrsim}
{\mathrel{\raisebox{-2.8pt}{\mbox{$\stackrel{\textstyle >}{\sim}$}}}}
\newcommand{\cxs}{c_{x,\sigma}}
\newcommand{\cus}{c_{u,\sigma}}
\newcommand{\axs}{a_{x,\sigma}}
\newcommand{\adxs}{a^\dagger_{x,\sigma}}
\newcommand{\bus}{b_{u,\sigma}}
\newcommand{\bdus}{b^\dagger_{u,\sigma}}

\newcommand{\Stot}{S_{\rm tot}}

\newcommand{\Smax}{S_{\rm max}}
\newcommand{\vac}{\Phi_{\rm vac}}

\newcommand{\la}{\lambda}

\newcommand{\Ne}{N_{\rm e}}

\renewcommand{\phi}{\varphi}
\newcommand{\ep}{\varepsilon}
\newcommand{\mG}{{\sf G}}
\newcommand{\Lao}{\La_{o}}
\newcommand{\Eo}{\calE_{o}}
\newcommand{\Io}{\calI_{o}}
\newcommand{\exfor}[1]{\backslash\{#1\}}
\newcommand{\Eup}{E_{\up}}
\newcommand{\Edn}{E_{\dn}}
\newcommand{\Iup}{I_{\up}}
\newcommand{\Idn}{I_{\dn}}
\newcommand{\Hfn}{{\cal H}_{\rm fin}^{(\nu)}}
\newcommand{\Pfn}{{\cal P}_{\rm fin}^{(\nu)}}
\newcommand{\hen}{h_{\rm eff}^{(\nu)}}
\newcommand{\htn}{\tilde{h}_{o}^{(\nu)}}
\newcommand{\Hfz}{{\cal H}_{\rm fin}^{(0)}}
\newcommand{\Pfz}{{\cal P}_{\rm fin}^{(0)}}
\newcommand{\hez}{h_{\rm eff}^{(0)}}
\newcommand{\htz}{\tilde{h}_{o}^{(0)}}
\newcommand{\Gz}{{\cal S}^{(0)}}
\newcommand{\Vz}{{\cal V}^{(0)}}
\newcommand{\Mz}{{\cal M}^{(0)}}
\newcommand{\Gn}{{\cal S}^{(\nu)}}
\newcommand{\hilb}{\mathfrak{h}}
\newcommand{\comp}{\mathbb{C}}
\newcommand{\real}{\mathbb{R}}
\newcommand{\inte}{\mathbb{Z}}

\newcommand{\vsigma}{\boldsymbol{\sigma}}

\makeatletter
\long\def\@makecaption#1#2{{\small
\advance\leftskip1cm
\advance\rightskip1cm
\vskip\abovecaptionskip
\sbox\@tempboxa{#1: #2}%
\ifdim \wd\@tempboxa >\hsize
 #1: #2\par
\else
\global \@minipagefalse
\hb@xt@\hsize{\hfil\box\@tempboxa\hfil}%
\fi
\vskip\belowcaptionskip}}
\makeatother
\begin{document}
\begin{flushright}
\footnotesize
to appear in Commun. Math. Phys.
\end{flushright}
\begin{center}
{\bf Ferromagnetism in the Hubbard model }
\par\bigskip
--- A constructive approach ---

\par\vfil
Hal Tasaki
\footnote{hal.tasaki\makeatletter @\makeatother gakushuin.ac.jp,
http://www.gakushuin.ac.jp/\( \tilde{\ } \)881791/
}
\par\bigskip
{\footnotesize\sl Department of Physics, Gakushuin University,
Tokyo 171-8588, JAPAN}
\end{center}
\par\vfil
\begin{abstract}
It is believed that strong ferromagnetic orders in some solids 
are generated by subtle interplay between quantum many-body effects 
and spin-independent Coulomb interactions between electrons.
Here we describe our rigorous and constructive approach
to ferromagnetism in the Hubbard model, which is 
a standard
idealized model for strongly interacting electrons in a solid.
We introduce a class of Hubbard models in any dimensions which are
nonsingular in the sense that both the Coulomb interaction and the
density of states (at the Fermi level) are finite.
We then prove that the ground states of the models exhibit
saturated ferromagnetism, i.e., have maximum total spins.
Combined with our earlier results,
the present work provides
nonsingular models of itinerant electrons 
with only spin-independent interactions
where low energy behaviors are proved to be
that of a ``healthy'' ferromagnetic insulator.
\end{abstract}
\newpage
\tableofcontents
\newpage
\section{Introduction}
\label{s:itro}

The origin of strong ferromagnetic order observed
in some solids has long been a mystery in physical
science.
After Heisenberg \cite{Heisenberg28}, 
it became clear that the ultimate origin of ferromagnetism
lies in a subtle interplay between quantum many-body
effects and strong Coulomb interaction between
electrons.
To provide convincing derivations of ferromagnetism
in concrete models of many electrons, however,
remained unsolved 
(even on a heuristic level)
for a long time.

The problem is difficult
because neither quantum many-body effects, nor
the Coulomb interaction alone favors ferromagnetism
(or any magnetic ordering).
One must deal with an interplay
of both the factors.
Moreover, the intrinsically nonperturbative nature
of the phenomenon makes the problem almost
impossible to attack within conventional
perturbative methods in condensed matter physics.
A generic many-electron system without interactions
is known to have a paramagnetic ground state,
a phenomenon known as Pauli paramagnetism.
In order to destabilize Pauli
paramagnetism and stabilize ferromagnetism,
one must have a sufficiently large interaction.
For example, a heuristic argument due to 
Stoner implies the criterion that 
\( U\,D_{\rm F}\gtrsim1 \) is necessary to stabilize
ferromagnetism, where \( U \) is the on-site
Coulomb interaction and \( D_{\rm F} \)
is the density of states at the Fermi level\footnote{
This is only a heuristic criterion, and there
are many counterexamples.
}.
This is the well-known ``competition''
between quantum dynamics and 
Coulomb interaction.

In the present paper,
we describe our constructive and
mathematically rigorous approach to the
origin of ferromagnetism.
This is a continuation of the series of works
\cite{92e,93d,localPRL,JSP},
and the main result of the present paper
was described in \cite{nonflat} for a
special one-dimensional model.
Here {\em we present
a class of Hubbard models in any dimensions
with a finite density of states 
(at the Fermi level)
and finite interactions, and prove that their
ground states are ferromagnetic.}\/
Combined with our earlier work \cite{localPRL,JSP},
this provides
{\em a class of nonsingular
models of itinerant electrons
(with only spin-independent interactions)
in which low energy behaviors
(i.e., the nature of the ground states
and the low-lying excitations)
are rigorously proved to be those expected in 
ferromagnetic insulators}\/.
We hope that the present work becomes
a starting point of further investigations
of deep interplay between quantum dynamics
and nonlinear interactions in strongly interacting
quantum many-body systems.

\bigskip

The present paper is written in a nearly
self-contained manner.
In Section~\ref{s:Hubdef}, we give the 
definition of the Hubbard model.
In Section~\ref{s:Rferro}, we briefly review
rigorous results about ferromagnetism
in the Hubbard model, and motivate the
present paper.
In Section~\ref{s:typ}, we summarize,
in a typical class of models,
the main results of our constructive program 
in the present and the previous works
of ours.
The reader who is interested in the new
physical results is invited to
start from this section.
In Section~\ref{s:main}, which is 
the main section of the paper,
we define our models in the most
general setting and state our conclusions
precisely.
Final section~\ref{s:proof} is devoted to the
proof of the main theorem.

\section{Definition of the Hubbard model}
\label{s:Hubdef}

The Hubbard model is a standard simple model of
interacting itinerant electrons in a solid.
Although this model is too idealized to be
regarded as a quantitatively reliable
model of real solids,
it contains physically essential
features of interacting
itinerant electron systems.
It is expected to exhibit various
phenomena including antiferromagnetism,
ferromagnetism, ferrimagnetism, 
superconductivity, and
metal-insulator transition.
Some (but not all) of these phenomena have been
treated rigorously in some cases
\cite{Liebreview}.

In the present section, we define
the Hubbard model in the general
setting, and fix our notation.
We leave details and backgrounds to more
careful reviews (such as \cite{Liebreview,PTP,JPC})
and present only the minimum necessary ingredients.

\subsection{Basic operators}
\label{s:op}

Let a lattice
\( \Lambda \) be a finite set whose
elements \( r,s,\cdots\in\Lambda \)
are called {\em sites}\/.
A site represents an atomic orbit in 
a solid.

For each \( r\in\Lambda \) and 
\( \sigma=\up,\dn \),
we define the creation and the annihilation
operators \( c^\dagger_{r,\sigma} \)
and \( c_{r,\sigma} \) 
for an electron at site \( r \) with
spin \( \sigma \).
These operators satisfy the 
canonical anticommutation relations
\begin{equation}
	\{c^\dagger_{r,\sigma},c_{s,\tau}\}
	=
	\delta_{r,s}\,\delta_{\sigma,\tau},
	\label{eq:cac}
\end{equation}
and 
\begin{equation}
	\{c^\dagger_{r,\sigma},c^\dagger_{s,\tau}\}
	=
	\{c_{r,\sigma},c_{s,\tau}\}
	=0,
	\label{eq:cac2}
\end{equation}
for any \( r,s\in\Lambda \)
and \( \sigma,\tau=\up,\dn \),
where \( \{A,B\}=AB+BA \).
The number operator is defined by
\begin{equation}
	n_{r,\sigma}=c^\dagger_{r,\sigma}\,c_{r,\sigma},
	\label{eq:ndef}
\end{equation}
which has eigenvalues 0 and 1.

The Hilbert space of the model is constructed
as follows.
Let \( \vac \) be a normalized
vector state which satisfies
\( c_{r,\sigma}\vac=0 \)
for any \( r\in\Lambda \) and 
\( \sigma=\up,\dn \).
Physically \( \vac \) corresponds to
a  state where there are no
electrons in the system.
Then for arbitrary subsets 
\( \Lambda_{\up},\Lambda_{\dn}\subset\Lambda \),
we define a state\footnote{
Throughout the present paper, we assume that the sites
in the lattice are ordered (in an arbitrary
but fixed manner),
and products of fermion operators
respect the ordering.
}
\begin{equation}
	\rbk{\prod_{r\in\Lambda_{\up}}
	c^\dagger_{r,\up}}
	\rbk{\prod_{r\in\Lambda_{\dn}}
	c^\dagger_{r,\dn}}
	\vac,
	\label{eq:basis}
\end{equation}
in which sites in \( \Lambda_{\up} \)
are occupied by up-spin electrons
and sites in \( \Lambda_{\dn} \)
by down-spin electrons.
The Hilbert space for the system
with \( \Ne \) electrons is spanned by
the basis states 
(\ref{eq:basis}) with all subsets
\( \La_{\up} \) and \( \La_{\dn} \)
such that\footnote{
Throughout the present paper
\( |S| \) denotes the number of elements
in a set \( S \).
} 
\( |\La_{\up}|+|\La_{\dn}|=\Ne \).

We finally define total spin operators 
\( \hat{\bf S}_{\rm tot}
= (\hat{S}^{(1)}_{\rm tot},
\hat{S}^{(2)}_{\rm tot},
\hat{S}^{(3)}_{\rm tot})\) by
\begin{equation}
	\hat{S}^{(\alpha)}_{\rm tot}
	=
	\frac{1}{2}
	\sumtwo{r\in\La}{\sigma,\tau=\up,\dn}
	c^\dagger_{r,\sigma}(p^{(\alpha)})_{\sigma,\tau}\,
	c_{r,\tau},
	\label{eq:Stot}
\end{equation}
for \( \alpha=1 \), 2, and 3.
Here \( p^{(\alpha)} \) are the Pauli matrices
defined by
\begin{equation}
	p^{(1)}=\pmatrix{0&1\cr1&0\cr},\quad
	p^{(2)}=\pmatrix{0&-i\cr i&0\cr},\quad
	p^{(3)}=\pmatrix{1&0\cr0&-1\cr}.
	\label{eq:Pauli}
\end{equation}
The operators \( \hat{\bf S}_{\rm tot} \)
are the generators of  
\( SU(2) \) rotations of the total 
spin angular momentum of the system.
As usual we denote the eigenvalue
of \( (\hat{\bf S}_{\rm tot})^2 \) as
\( S_{\rm tot}(S_{\rm tot}+1) \).
The maximum possible value of 
\( S_{\rm tot} \)
is \( \Ne/2 \) when \( \Ne\le|\La| \).

\subsection{General Hamiltonian}
\label{s:genHam}
The model is characterized by the
hopping amplitudes 
\( t_{r,s}=t_{s,r}\in\real \)
defined for all \( r,s\in\La \),
and the magnitude \( U>0 \) of the
on-site Coulomb interaction.
Physically, \( t_{r,s} \) represents
the quantum mechanical amplitude for
an electron to hop from the site
\( s \) to site \( r \) when \( s\ne r \),
and the on-site potential when \( r=s \).
Usually \( t_{r,s} \) is non-negligible
only when the two sites \( r \) and \( s \)
are close to each other.

We then define the general Hubbard Hamiltonian
as
\begin{equation}
	H=\sumtwo{r,s\in\La}{\sigma=\up,\dn}
	t_{r,s}\,c^\dagger_{r,\sigma}c_{s,\sigma}
	+
	U\sum_{r\in\La}n_{r,\up}n_{r,\dn}.
	\label{eq:genH}
\end{equation}
Here the first term describes quantum mechanical
motion of electrons which hop around the lattice
according to the amplitude \( t_{r,s} \).

The second term represents nonlinear interactions
between electrons.
There is an increase in energy by
\( U>0 \) for each doubly occupied site, i.e.,
a site which is occupied by both up-spin electron
and down-spin electron.
This is a highly idealized treatment of 
the Coulomb interaction between electrons.

The Hamiltonian which consists
only of the first term in (\ref{eq:genH})
describes the free tight-binding electron model.
It is not very difficult to analyze
this model
especially when the hopping amplitude
\( t_{r,s} \) has a translation invariance.
The Hamiltonian can be diagonalized in the
states in which electrons behave as ``waves.''

The Hamiltonian which consists
only of the second term in (\ref{eq:genH})
is also easy to study.
The Hamiltonian is already diagonalized
in the basis states (\ref{eq:basis}),
in which electrons behave as
``particles.''

When both the first and the second terms in
(\ref{eq:genH}) are present,
a ``competition'' between wave-like nature
and particle-like nature of electrons
takes place.
The competition generates rich 
nontrivial phenomena including ferromagnetism.
To investigate these phenomena is a main motivation
in the study of the Hubbard model.

\section{Rigorous results
about ferromagnetism in the Hubbard model}
\label{s:Rferro}
In the present section, we formulate
the problem of saturated ferromagnetism in the
Hubbard model.
We then give a brief review of the
rigorous results about ferromagnetism in the Hubbard model,
and explain background of the present work.
For more careful reviews,
see \cite{PTP,JPC}.

\subsection{Saturated ferromagnetism in
the ground states}
\label{s:ferro}
It is easily shown that the Hamiltonian 
(\ref{eq:genH}) commutes with the total
spin operators 
\( \hat{S}^{(\alpha)}_{\rm tot} \).
Therefore one can look for
simultaneous eigenstates of 
\( H \) and \( (\hat{\bf S}_{\rm tot})^2 \).

When all the ground states of 
the Hamiltonian \( H \)
(with a fixed electron number
\( \Ne\le|\La| \)) are eigenstate of
\( (\hat{\bf S}_{\rm tot})^2 \)
with \( S_{\rm tot}=\Ne/2 \),
we say that the model exhibits
{\em saturated ferromagnetism}\/.
This is the strongest form of ferromagnetism
since \( \Ne/2 \) is the maximum possible 
value for \( S_{\rm tot} \).

\subsection{Ferromagnetism of Nagaoka and Thouless}
\label{s:Nagaoka}

The first rigorous and nontrivial result
about saturated ferromagnetism in the Hubbard
model is due to Nagaoka \cite{Nagaoka66} and to Thouless
\cite{Thouless65}.
It was proved that the Hubbard model on
a class of lattices (which includes
most of the standard lattices
in two and three dimensions)
with \( t_{r,s}\ge0 \)
exhibits saturated ferromagnetism
when \( \Ne=|\La|-1 \)
and \( U=\infty \).
In other words the model is not
allowed to have any doubly occupied sites,
and there is only one site without
an electron.

The ferromagnetism of Nagaoka and
Thouless is quite important since it showed
for the first time that the Hubbard model can generate
ferromagnetism through nontrivial
interplay between quantum dynamics and Coulomb
interaction.
Subsequent studies, however, have suggested that their
mechanism of ferromagnetism is restricted
to special situation with infinite \( U \)
and a single hole.
See Section~4 of \cite{PTP} for a modern proof
and further discussions.

\subsection{Lieb's ferrimagnetism
and flat-band ferromagnetism}
\label{s:flat}

\begin{figure}
\centerline{\epsfig{file=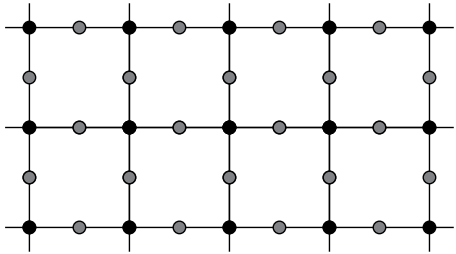,width=7cm}}
\caption[dummy]{
The so called copper oxide lattice.
As a consequence of Lieb's theorem \cite{Lieb89},
it is proved that the Hubbard model with \( U>0 \)
on this lattice has \( S_{\rm tot}=\Ne/6 \)
when \( \Ne=|\La| \).
}
\label{f:CuO}
\end{figure}

In 1989, after more than two decades from the
works of Nagaoka and Thouless,
Lieb proved an important theorem for the Hubbard model
with \( \Ne=|\La| \) (i.e., half-filling)
on a bipartite lattice \cite{Lieb89}.
For the Hubbard model with \( U>0 \)
on lattices which have two sublattices
with different numbers of sites,
Lieb's theorem implies the existence
of ferrimagnetism, a weaker version of 
ferromagnetism.
A typical example is the Hubbard model 
on the so called copper oxide lattice
of Fig.~\ref{f:CuO},
where the ground states are proved to have
\( \Stot=\Ne/6 \) when \( \Ne=|\La| \).
The models 
exhibiting Lieb's ferrimagnetism
have peculiar single-electron
band structures where the band at the middle
of the spectrum is completely flat
(or dispersionless).
One may regard Lieb's ferrimagnetism
as a precursor to the flat-band ferromagnetism
that we shall discuss.

\begin{figure}
\centerline{\epsfig{file=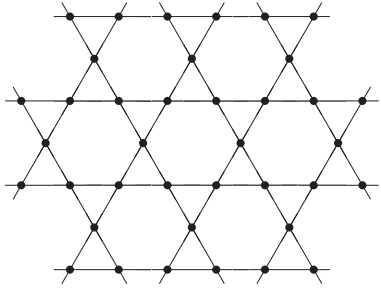,width=5cm}}
\caption[dummy]{
The kagom\'{e} lattice is the line graph of the hexagonal lattice.
Mielke \cite{Mielke91a,Mielke91b,Mielke92}
showed that the Hubbard model on the kagom\'{e} lattice 
exhibits ferromagnetism when \( \Ne=|\La|/3 \)
for any \( U>0 \).
}
\label{f:kagome}
\end{figure}

Flat-band ferromagnetism was discovered 
first by Mielke \cite{Mielke91a,Mielke91b,Mielke92}
and then by Tasaki \cite{92e,93d}.
Mielke treated the Hubbard model on a general line
graph, where \( t_{r,s}=t>0 \) for 
those pairs \( (r,s) \) corresponding to the edges (or bonds) of the
lattice, and \( t_{r,s}=0 \)
otherwise.
The models have peculiar band structure where the 
lowest single-electron band is completely flat.
Mielke
proved that the models 
with \( U>0 \) exhibit saturated ferromagnetism
for suitable electron numbers
which correspond to the half-filling of the lowest bands.
A typical example (and the most beautiful
example of flat-band ferromagnetism)
is the Hubbard model on the kagom\'{e} lattice
of Fig.\ref{f:kagome},
which was proved to exhibit saturated ferromagnetism
when \( \Ne=|\La|/3 \).
See also \cite{PF1,PF2,PF3} for Mielke's results
on Hubbard models with partially flat bands.

\begin{figure}
\centerline{\epsfig{file=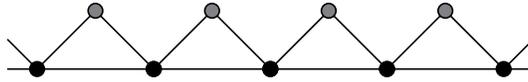,width=7cm}}
\caption[dummy]{
Tasaki's flat-band Hubbard model in one dimension
\cite{92e,93d}.
The hopping amplitude \( t_{r,s} \) is \( \nu^2t \)
for the horizontal bonds and
\( \nu t \) for the diagonal bonds.
The sites in the upper and the lower rows have
on-site potential \( t_{r,r} \)
which equal \( t \) and \( 2\nu^2t \), respectively.
When \( \Ne=|\La|/2 \), the model exhibits
saturated ferromagnetism for
any \( t>0 \), \( \nu>0 \),
and \( U>0 \).
See Theorem~\ref{t:flat}.
}
\label{f:1Dflat}
\end{figure}

Tasaki \cite{92e,93d}
proposed his version of Hubbard models
with flat lowest bands, and proved the
existence of saturated ferromagnetism for \( U>0 \)
when the lowest bands are half-filled.
As can be seen from the one-dimensional
example in Fig.~\ref{f:1Dflat},
his models have two different kinds of
lattice sites which are sometimes
interpreted as metallic and oxide atoms,
and have next nearest neighbor hopping amplitudes.
By fine-tuning the hopping amplitudes
and the on-site potentials, 
the lowest band becomes flat.
See \cite{Sekizawa} for an extension.

A common feature of Lieb's ferrimagnetism
and Mielke's and Tasaki's ferromagnetism
is that 
their models have single-electron
bands which are totally flat (i.e., dispersionless),
and the magnetization is supported by
electrons in the flat bands.
(For Lieb's ferrimagnetism, the latter statement
is correct only in a vague sense.)
This observation is consistent with the 
Stoner criterion which states that 
large \( U D_{\rm F} \) favors ferromagnetism.
Here the criterion is realized by infinitely large
density of states 
\( D_{\rm F} \).

The works of Lieb, Mielke, and Tasaki have shown
that rich classes of Hubbard models on 
slightly complicated lattices exhibit
nontrivial magnetic behavior.
Such a view may be helpful in understanding 
insulating ferromagnetism observed
in a cuprate \cite{MizunoMasudaHirabayashi93,PTP},
and has even motivated some scientists
to design novel ferromagnetic materials.
See \cite{ASKA} and references therein.
But one should not forget that the Hubbard model is a
highly idealized model.
To find implications of the results for the Hubbard model
in realistic many-electron systems defined in continuum space
is a formidably difficult but a challenging problem.
See, for example, \cite{ASKA,KM}.

\subsection{Beyond flat-band ferromagnetism}
\label{s:beyond}

Although Lieb's ferrimagnetism and 
Mielke's and Tasaki's flat-band ferromagnetism
certainly have shed novel
light on the mechanisms of magnetic ordering
in interacting electron systems,
they do not deal with the true ``competition''
between quantum dynamics and Coulomb interactions.
When the Coulomb interaction \( U \) is vanishing,
all of their models have highly degenerate ground
states.
The degeneracy reflects the existence of
completely flat bands.
Among these degenerate ground states for \( U=0 \),
there are ferrimagnetic or ferromagnetic
states as well as states with much smaller
magnetization.
The role of the Coulomb interaction in these models
is simply to lift the huge degeneracy
and ``select'' the states with highest magnetization
as unique ground states.
Consequently ferrimagnetism or ferromagnetism
in these models
takes place for any values of \( U>0 \).
In other words magnetic ordering is stabilized 
by infinitesimally small Coulomb interaction.
This is quite different from situations in realistic systems
where the interaction 
must be greater than some positive critical value in order
to destabilize
Pauli paramagnetism and get magnetic ordering.

It may be needless to say that the 
existence of completely flat lowest bands
(especially in Tasaki's models) is
unrealistic, or even pathological.
The flatness of the bands is destroyed
by arbitrarily small generic perturbation,
and is far from robust.

It was therefore highly desirable to go beyond
flat band models.
A natural step was to modify the model
by adding extra hopping terms to 
the Hamiltonian thus making the flat
band dispersive, and then to show that the
magnetic ordering survives.
One can only hope this scenario to work
for sufficiently large \( U \)
since magnetic ordering becomes
truly a nonperturbative phenomenon
when the band is not flat.

\begin{figure}
\centerline{\epsfig{file=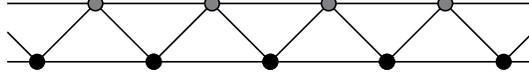,width=7cm}}
\caption[dummy]{
Tasaki's nearly-flat-band Hubbard model in one dimension
\cite{nonflat}.
The hopping amplitude \( t_{r,s} \) is \( -\nu^2s \)
and \( \nu^2t \)
for the horizontal bonds 
in the upper and the lower rows, respectively, 
and
\( \nu (t+s) \) for the diagonal bonds.
The sites in the upper and the lower row have
on-site potential \( t_{r,r} \)
which equal \( t-2\nu^2s \) and \( 2\nu^2t-s \), respectively.
The model has two bands which are not flat.
When \( \Ne=|\La|/2 \), the model exhibits
saturated ferromagnetism for
sufficiently large \( U/s \) and \( t/s \)
for any \( \nu>0 \).
When \( \nu=1/\sqrt{2} \), for example,
the appearance of ferromagnetism is proved
for \( t/s\ge1.6 \) and sufficiently large \( U/s \).
See Theorem~\ref{t:non-f}.
}
\label{f:1Dnf}
\end{figure}

As the first step in this direction,
the local stability of the ferromagnetic
state was proved in models obtained by 
adding arbitrary small short-range hopping terms 
to Tasaki's version of flat-band Hubbard models \cite{localPRL,JSP}.
In this work, it was also shown that low-lying excitation energy
above the ferromagnetic state has the dispersion relation
expected for a magnon excitation.
Then it was proved in \cite{nonflat} that a 
one-dimensional 
Hubbard model with non-flat bands exhibits saturated 
ferromagnetism for sufficiently large \( U \).
The model was obtained by adding extra nearest neighbor
hopping terms to Tasaki's one-dimensional flat-band Hubbard model
as in Fig.~\ref{f:1Dnf}.
This was the first rigorous example of ferromagnetism
in an electron system without any singularities,
i.e., with finite interaction and finite density of states.
Shen \cite{Shen98}
announced a computer assisted extension of
the proof in \cite{nonflat} 
to some higher dimensional models.
The method in \cite{nonflat} 
inspired similar rigorous works in different
classes of 
Hubbard models \cite{TI1,TI2}.
In particular Tanaka and Ueda \cite{Tanakanew} 
recently
succeeded in proving the existence
of saturated ferromagnetism in a Hubbard model
obtained by adding extra hopping terms to 
Mielke's flat band Hubbard model on the kagom\'{e}
lattice.
For closely related heuristic works,
see \cite{KusakabeAoki94b,PencShibaMilaTsukagoshi96}
and other references in Section~6.6 of \cite{PTP}.

The present work is an extension of that
in \cite{nonflat}.
We extend the theorem in \cite{nonflat}
to general models in higher dimensions.
As was noted in \cite{nonflat}, a straightforward extension of the method
in \cite{nonflat} applies to a class of higher dimensional models.
Instead of using such a method,
we here present a much more general
and simplified proof which naturally covers
a more general class of models.

\section{Ferromagnetism in typical
$d$-dimensional nearly-flat-band models}
\label{s:typ}

In the present section, we concentrate on a simple
class of models defined on decorated hypercubic lattices,
and precisely describe the 
results of the present paper and our previous
works.
Although our works cover much more general 
models, it may be useful for the readers to see 
what has been achieved in the context of 
simple models.
In short, we start from a concretely defined
non-singular model of itinerant electrons, and
prove that its low energy properties coincide
with what one expects in a ``healthy'' ferromagnet.

\begin{figure}
\centerline{\epsfig{file=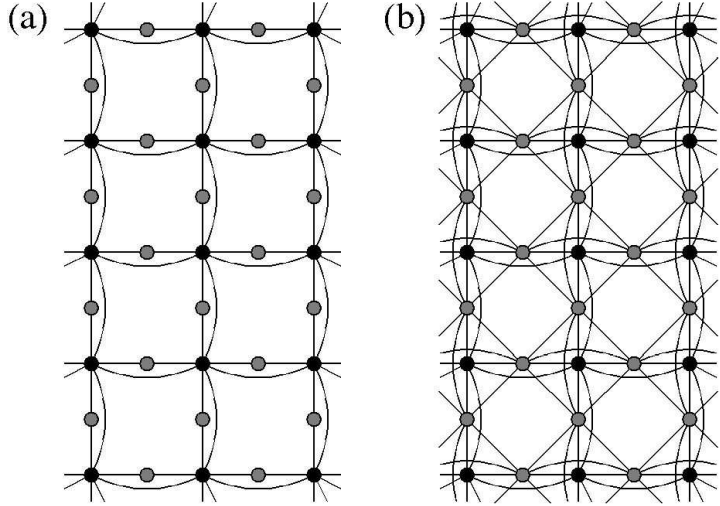,width=10cm}}
\caption[dummy]{
The lattice structure and the hopping amplitudes in
the two dimensional model.
The black dots are sites in \( \calE \),
and the gray dots are sites in \( \calI \).
(a) shows the flat-band model with \( s=0 \),
and (b) shows general model with \( s>0 \).
}
\label{f:2Dlattice}
\end{figure}

Let \( \calE \) denote (only in the present section)
the \( d \)-dimensional \( L\times\cdots\times L \) hypercubic
lattice with the unit lattice spacing and periodic
boundary conditions.
We let \( L>0 \) to be an odd integer.
We take a new site in the middle of each bond
(i.e., a pair of neighboring sites) in \( \calE \),
and denote by \( \calI \) (again only in this section)
the collection of all such sites.
We shall study the decorated hypercubic lattice
\( \La=\calE\cup\calI \) in the present section.
See Fig.~\ref{f:2Dlattice}.

We define a Hubbard model on \( \La \)
which is characterized by four 
parameters \( t>0 \), \( s>0 \),
\( \nu>0 \), and \( U>0 \).
The hopping amplitude of the model
is given by
\begin{equation}
	t_{r,s}=\cases{
	\nu(t+s)&
	if \( |r-s|=1/2 \);\cr
	\nu^2t&
	if \( r,s\in\calE \) and \( |r-s|=1 \);\cr
	-\nu^2s&
	if \( (r,s)\in\calB \);\cr
	2d\nu^2t-s&
	if \( r=s\in\calE \);\cr
	t-2\nu^2s&
	if \( r=s\in\calI \);\cr
	0&
	otherwise,
	}
	\label{eq:ttrs}
\end{equation}
where we set
\begin{equation}
	\calB=
	\set{(r,s)}{r,s\in\calI, \ |r-s|=1/\sqrt{2}}
	\cup
	\set{(r,s)}{r,s\in\calI,\  |r-s|=1, \ (r+s)/2\in\calE}.
	\label{eq:Btil}
\end{equation}
There are nearest neighbor and next nearest neighbor hopping amplitudes.
See Fig.~\ref{f:2Dlattice}.
This rather complicated expression for \( t_{r,s} \)
comes from a simple construction in Section~\ref{s:defmain}.
See (\ref{eq:H}).

\begin{figure}
\centerline{\epsfig{file=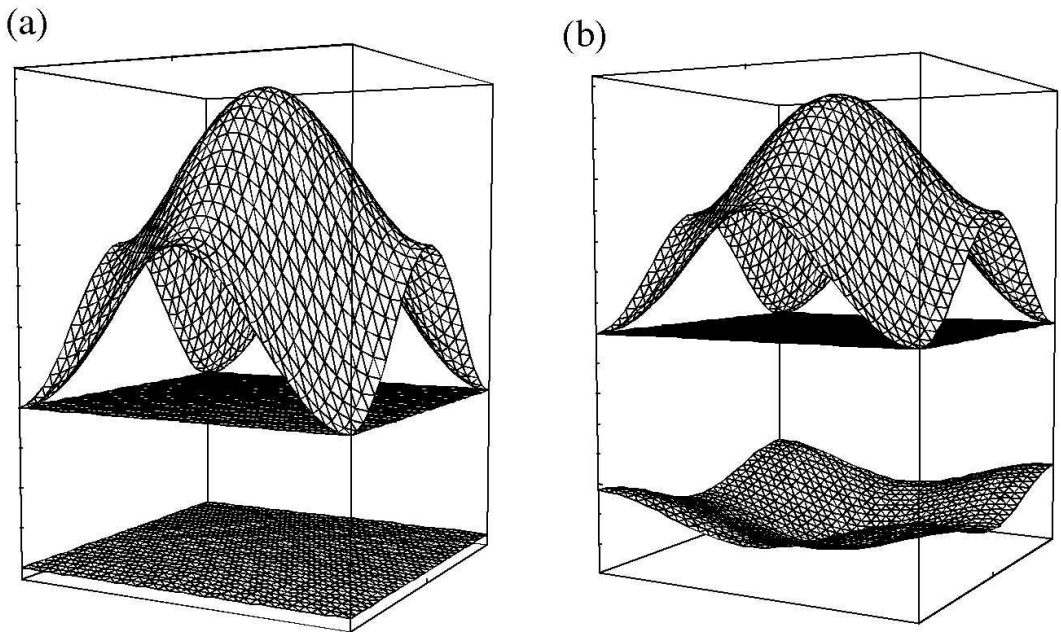,width=14cm}}
\caption[dummy]{
The dispersion relation (\ref{eq:ejk})
of the single-electron bands in
the two dimensional model.
The horizontal axes represent
\( -\pi\le k_{1},k_{2}\le\pi \),
and the vertical axis denotes the single-electron
energy.
(a) shows the flat-band model with \( s=0 \),
and (b) shows general model with \( s>0 \).
}
\label{f:2Dbands}
\end{figure}

One can easily calculate the single-electron
properties corresponding to the above
hopping amplitudes.
There are \( (d+1) \) single-electron
bands\footnote{
The readers unfamiliar with the notion of
bands may ignore this part or refer to
Appendix~E of \cite{PTP}.
} and their dispersion relations are
given by\footnote{
In our models,  all the bands have simple cosine
dispersion relations.
This is not the case in general multi-band systems, and
reflects a special character of our hopping amplitudes.
In this sense, our models may be regarded as a kind of
``idealized tight-binding models.''
Whether such models are useful in studying problems other than
ferromagnetism is an open question.
}
\begin{equation}
	\ep_{j}(k)=\cases{
	-s-2\nu^2s\sum_{\mu=1}^{d}(1+\cos k_{\mu})
	&if \( j=1 \);\cr
	t&if \( j=2,\cdots,d \);\cr
	t+2\nu^2t\sum_{\mu=1}^{d}(1+\cos k_{\mu})
	&if \( j=d+1 \).
	}
	\label{eq:ejk}
\end{equation}
Here \( k=(k_{1},\ldots,k_{d}) \)
is the wave vector in the set
\begin{equation}
	\calK=
	\set{(\frac{2\pi}{L}n_{1},\ldots,\frac{2\pi}{L}n_{d})}
	{n_{i}=0,\pm1,\pm2,\ldots,\pm\frac{L-1}{2}}.
	\label{eq:K}
\end{equation}
In the flat-band model with \( s=0 \),
all the bands except the uppermost band
with \( j=d+1 \) are dispersionless
(or flat) as in Fig.~\ref{f:2Dbands}~(a).
In a general model with \( s>0 \),
the lowest band becomes dispersive
as in Fig.~\ref{f:2Dbands}~(b).
Since our ferromagnetism is supported by electrons
in the lowest band, it is crucial that the
lowest band becomes dispersive.
Reflecting the special geometry of the decorated
lattice,
the intermediate bands with \( j=2,\ldots,d \)
are always dispersionless.
This, however, is not crucial to low energy
behavior of our model.
Indeed it is not difficult to add proper extra hopping terms
to the model to make all the bands dispersive
while maintaining the existence of
ferromagnetism.
See Section~\ref{s:ext}.

We consider the Hubbard model with the Hamiltonian
(\ref{eq:genH}),
the hopping amplitudes (\ref{eq:ttrs}),
and the electron number 
\( \Ne=|\calE|=L^d \).

We first recall the result about the flat-band
ferromagnetism proved in \cite{92e,93d}.
(See Section 6 of \cite{PTP} for the most
compact proof.)

\begin{theorem}[Flat-band ferromagnetism]
\label{t:flat}
Let \( s=0 \).
Then for arbitrary \( t>0 \), \( \nu>0 \),
and \( U>0 \),
the above model exhibits saturated ferromagnetism.
\end{theorem}

As we have stressed in Section~\ref{s:beyond},
ferromagnetism takes place for any positive
values of \( U \) in the flat-band models.
When the lowest band is no longer flat,
saturated ferromagnetism cannot take place for 
too small values of \( U>0 \).
This fact can be seen, for example,
from the following (easy and well-known)
theorem.
(See Section~3.3 of \cite{PTP} for a proof.)

\begin{theorem}[Instability of saturated ferromagnetism]
\label{t:non-f}
Let \( s>0 \) and  \( U<4\nu^2s \).
Then the lowest energy among the states with 
\( \Stot=\Smax-1 \) is strictly lower than the
lowest  energy among the states with 
\( \Stot=\Smax \).
This means that the ground state of the model
has \( \Stot<\Smax \), and hence the model does
not exhibit saturated ferromagnetism.
\end{theorem}

The theorem, unfortunately, does not tell us
what the ground states look like for small \( U \).
(We nevertheless believe that the model has ground
states with \( \Stot=0 \) for sufficiently small \( U \).)
It assures us, however, that the appearance
of saturated ferromagnetism, which is
established in the following theorem,
is a purely nonperturbative phenomenon.

\begin{theorem}[Ferromagnetism in nearly-flat-band models]
\label{t:near}
When \( t/s \), \( U/s \), and \( 1/\nu \)
are sufficiently large
(how large these quantities should be depend only on the
dimensionality \( d \)),
the above model exhibits saturated ferromagnetism.
\end{theorem}

This is a special case of our main theorem in the
present paper, Theorem~\ref{t:main}.
For the model with \( d=1 \),
one can prove the same statement for
any values of \( \nu>0 \).
See Section~\ref{s:ext}.
A computer assisted proof of the above
theorem for \( d=2,3 \)
(which makes use of an
extension of the method in \cite{nonflat})
was announced by Shen \cite{Shen98}.

Moreover our earlier results
in \cite{localPRL,JSP} about low-lying 
excitations also apply to the present model.
For any \( x\in{\inte}^d \),
define the translation operator
\( T_{x} \) by
\begin{equation}
	T_{x}\sqbk{
	\rbk{\prod_{r\in\Lambda_{\up}}
	c^\dagger_{r,\up}}
	\rbk{\prod_{r\in\Lambda_{\dn}}
	c^\dagger_{r,\dn}}
	\vac
	}
	=
	\rbk{\prod_{r\in\Lambda_{\up}}
	c^\dagger_{r+x,\up}}
	\rbk{\prod_{r\in\Lambda_{\dn}}
	c^\dagger_{r+x,\dn}}
	\vac,
	\label{eq:Tx}
\end{equation}
where we use periodic boundary conditions to identify
\( r+x \) with a site in \( \La \) (if necessary).
Then, for any \( k\in\calK \),
we let
\( E_{\rm SW}(k) \) be
the lowest possible energy among the states
that satisfy 
\( \hat{S}_{\rm tot}^{(3)}\Phi
=\{(\Ne/2)-1\}\Phi\)
and 
\( T_{x}[\Phi]=e^{ik\cdot x}\Phi \)
for any \( x \).
In other words, 
\( E_{\rm SW}(k) \)
is the lowest energy among the states
where a single spin is flipped
(from the ferromagnetic ground state)
and the total momentum is \( k \).
Then we have the following theorem.
(For more precise statements, see
Section~3.3 of \cite{JSP}.)

\begin{theorem}[Dispersion relation of low-lying
excitations]
\label{t:SW}
Let \( E_{\rm GS} \) be the ground state energy.
When \( t/s \), \( U/s \), \( t/U \), and \( 1/\nu \)
are sufficiently large,
one has
\begin{equation}
	F_{1}\,{4\nu^4U}\sum_{\mu=1}^d(\sin\frac{k_{\mu}}{2})^2
	\ge
	E_{\rm SW}(k)-E_{\rm GS}
	\ge
	F_{2}\,{4\nu^4U}\sum_{\mu=1}^d(\sin\frac{k_{\mu}}{2})^2,
	\label{eq:SW}
\end{equation}
for any \( k\in\calK \).
Moreover the constants \( F_{1} \)
and \( F_{2} \) tend to 1 as
\( s\to0 \) and \( \nu\to0 \).
\end{theorem}

Therefore, for sufficiently small \( s \) and \( \nu \),
we have an almost precise estimate
\begin{equation}
	E_{\rm SW}(k)-E_{\rm GS}
	\simeq
	{4\nu^4U}\sum_{\mu=1}^d(\sin\frac{k_{\mu}}{2})^2,
	\label{eq:SW2}
\end{equation}
about the low-lying excitation energies.
We note that this dispersion relation is what one expects
for the elementary magnon excitation in a 
ferromagnetic Heisenberg model on \( \calE \)
with the exchange interaction
\( J_{\rm eff}=2\nu^4U \).

To summarize, we have obtained a class of non-singular models
of itinerant electrons\footnote{
It is true that the Hubbard model itself is ``singular''
when compared with more realistic models in continuum.
But this is a consequence of the  way of describing physical
systems, and does not necessarily mean that underlying system 
(if any) is singular.
We believe, on the other hand, that the models with
\( U=\infty \) or \( D_{\rm F}=\infty \) have
more manifest singularities.
} (with only spin-independent interactions)
whose low energy behaviors
are rigorously proved to be that of a ``healthy''
insulating ferromagnet\footnote{
It should be noted that insulating ferromagnets are rather
rare in reality.
To prove the existence of metallic ferromagnetism,
in which a set of electrons contribute both to 
conduction and magnetism, 
in certain version of the Hubbard model
is a challenging open problem \cite{PTP}.
}.
By a ``healthy'' insulator, we mean an itinerant electron
system whose low energy properties can effectively
be described by an appropriate quantum spin systems.
Although we can hardly claim that our model is realistic,
the similarity with ferromagnetism observed in a
cuprate (see Section~7.1 of \cite{PTP})
suggests that our models share some features 
with some of the existing ferromagnetic insulators.

Let us finally discuss whether our ferromagnetism is 
robust against perturbations.
We note that Theorem~\ref{t:SW} about the low-lying
excitation is still valid when one adds small arbitrary translation 
invariant perturbation to the hopping amplitudes\footnote{
One must also replace \( E_{\rm GS} \) in the 
theorem with the lowest energy of
the states with \( \Stot=\Smax \).
}.
In other words, local stability of the ferromagnetic state
is proved for slightly perturbed models.
Since it is generally believed that local stability of ferromagnetism
implies global stability (see \cite{PF2} for a related rigorous result),
this strongly suggests that the global stability of ferromagnetism
(as is stated in Theorem~\ref{t:near}) is valid for general
perturbed models.

\section{The model and main results}
\label{s:main}
\subsection{Construction of the lattice}
\label{s:lattice}
We define our lattice and the Hubbard model on it.

Let us give a brief explanation first.
Our lattice \( \La \) consists of two kinds of sites called 
external sites and internal sites.
The sets of all the external and the internal sites are denoted
as \( \calE \) and \( \calI \), respectively.
In the model of Fig.~\ref{f:2Dlattice}, for example,
the black dots are external sites and gray dots are
internal sites.
The whole lattice \( \La \) is decomposed into
a union of overlapping cells.
Each cell contains a single internal site
and \( n \) (\( n\ge2 \)) external sites.
An internal site \( u \) belongs to exactly one
cell (denoted as \( C_u \)), while 
an external site \( x \)  belongs to 
\( m \) (\( m\ge2 \)) cells.
In Fig.~\ref{f:2Dlattice}, a bond which consists
of two black dots and a gray dot is a cell.

To be more precise,
let us define the general lattice by
using the ``cell construction''
as in \cite{PTP}.
This allows us to cover a general class of models in
a unified manner.

We fix two integers \( n,m\ge2 \)
which will characterize our lattice.
Let the basic cell be a set of 
\( (n+1) \) sites written as
\begin{equation}
	C=\{u,x_{1},x_{2},\cdots,x_{n}\}.
	\label{eq:Cell}
\end{equation}
For convenience, 
we call \( u \) the {\em internal site}\/ of \( C \),
and \( x_{1},x_{2},\cdots,x_{n} \)
the {\em external sites}\/.

To form the lattice \( \La \),
we assemble \( M \) identical copies of
the basic cell, and
identify external sites from \( m \)
distinct cells regarding them as a single
site.
We do not make such identifications for
internal sites.
We assume that the lattice \( \La \)
thus constructed is connected.
Usually \( \Lambda \) becomes a periodic lattice
by this construction.

The lattice \( \La \) is naturally 
decomposed as
\begin{equation}
	\La=\calI\cup\calE,
	\label{eq:LIE}
\end{equation}
where \( \calI \) and \( \calE \) are the sets
of internal sites and external sites,
respectively.
From the above construction, we see that the
numbers of sites in these sublattices are
\( |\calI|=M \) and 
\( |\calE|=nM/m \).

\begin{figure}
\centerline{\epsfig{file=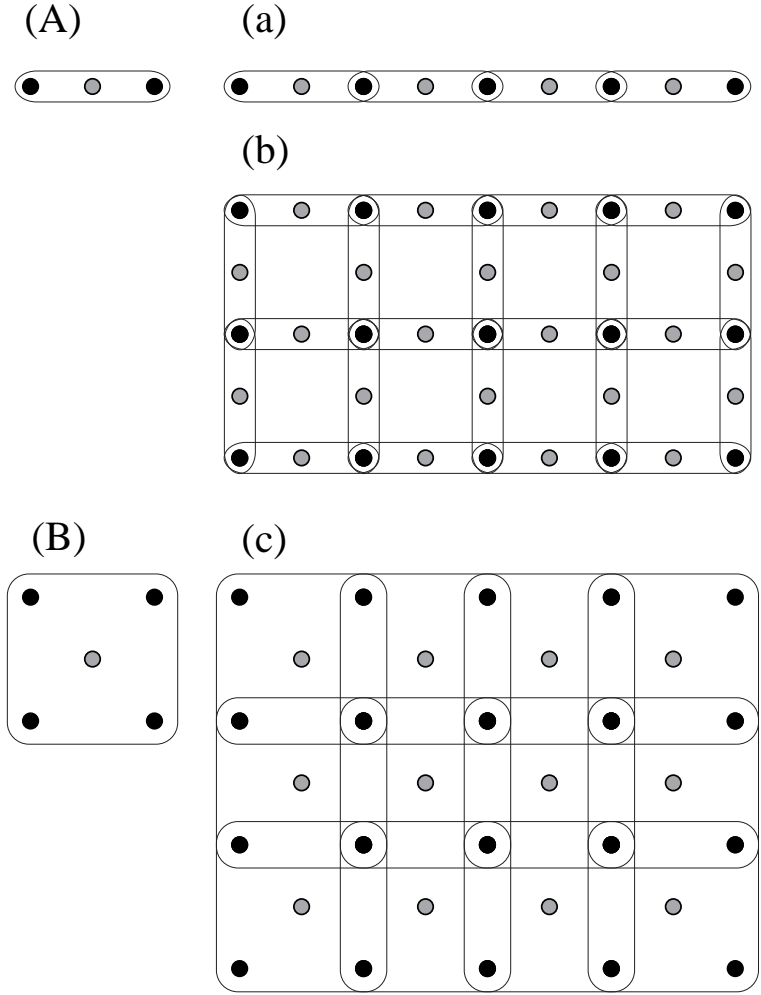,width=10cm}}
\caption[dummy]{
Examples of cells and lattices.
The black dots represent external sites,
and the gray dots represent internal sites.
 (A)~From the 
cell with three sites (\( n=2 \)), 
one can form (a)~a one-dimensional
lattice (which is drawn as the delta chain or
the zigzag chain in Figs.~\ref{f:1Dflat} and \ref{f:1Dnf}) by 
identifying two external sites (\( m=2 \)), or (b)~a decorated square lattice
(which also appears in Fig.~\ref{f:2Dlattice}) by 
identifying four (\( m=4 \)).
(B)~From the cell with five sites (\( n=4 \)), one can form,
for example, (c)~another 
decorated square lattice
(which will appear in Fig.~\ref{f:another})
by identifying four external sites (\( m=4 \)).
There are many similar examples in higher dimensions.
}
\label{f:lattices}
\end{figure}

See Figure~\ref{f:lattices} for some examples of the 
basic cell and corresponding lattices.
The examples treated in Section~\ref{s:typ}
are obtained by considering the cell with
\( n=2 \), and setting \( m=2d \).

We can easily treat models where \( n \)
and \( m \) are not identical for different
cells, but we here concentrate on the
simplest case with constant \( n \)
and \( m \).
(We still can treat a variety of lattices
by choosing different \( n \), \( m \),
and ways of assembling the cells.)

For an internal site \( u\in\calI \),
we denote by \( C_{u}\subset\La \)
the cell which contains the site \( u \).
For an external site \( x\in\calE \),
we denote by \( \La_{x}\subset\La \)
the union of \( m \) cells which contain
the site \( x \).

\subsection{Fermion operators}
\label{s:Fop}
We define special fermion operators 
which will be crucial for our analysis.
Let \( \nu>0 \) be a constant.
(We note that \( 1/\nu \) corresponds to \( \la \) in our previous
publications \cite{92e,93d,localPRL,JSP,PTP}.)
For \( x\in\calE \), let
\begin{equation}
	\axs=\cxs-
	\nu\sum_{u\in\La_{x}\cap\calI}\cus,
	\label{eq:axs}
\end{equation}
where the sum is over \( m \)
internal sites adjacent to \( x \).
Similarly for \( u\in\calI \), let
\begin{equation}
	\bus=\cus+
	\nu\sum_{x\in C_{u}\exfor{u}}
	\cxs,
	\label{eq:bus}
\end{equation}
where the sum is over the \( n \) external sites
adjacent to \( u \).

From the anticommutation relations 
(\ref{eq:cac}) for the
basic \( c \) operators, one can easily verify that
\begin{equation}
	\{a^\dagger_{x,\sigma},b_{u,\tau}\}
	=0
	\label{eq:ab}
\end{equation}
for any \( x\in\calE \), \( u\in\calI \),
and \( \sigma,\tau=\up,\dn \).
This means that the \( a \) operators and
the \( b \) operators correspond to 
distinct spaces of electrons.
We shall discuss more about this point
in Section~\ref{s:band}.

The anticommutation relations between
the \( a \) operators are
\begin{equation}
	\{\adxs,a_{y,\tau}\}=
	\cases{
	1+m\nu^2,&if \( x=y \), \( \sigma=\tau \);\cr
	\ell_{x,y}\,\nu^2,&if \( x\ne y \), \( \sigma=\tau \);\cr
	0,&if \( \sigma\ne\tau \).
	}
	\label{eq:aa}
\end{equation}
For \( x,y\in\calE \), we defined
\begin{equation}
	\ell_{x,y}=|\La_{x}\cap\La_{y}\cap\calI|,
	\label{eq:lxy}
\end{equation}
which is the
number of distinct cells which contain
both \( x \) and \( y \).
For the \( b \) operators, we similarly have
\begin{equation}
	\{\bdus,b_{v,\tau}\}=
	\cases{
	1+n\nu^2,&if \( u=v \), \( \sigma=\tau \);\cr
	\ell_{u,v}\,\nu^2,&if \( u\ne v \), \( \sigma=\tau \);\cr
	0,&if \( \sigma\ne\tau \).
	}
	\label{eq:bb}
\end{equation}
For \( u,v\in\calI \), we defined
\begin{equation}
	\ell_{u,v}=|C_{u}\cap C_{v}\cap\calE|,
	\label{eq:luv}
\end{equation}
which is
the number of external sites
which are adjacent to both \( u \) and \( v \).
One sees that \( a \) operators or \( b \) operators
simply anticommute with each other if the
reference sites are sufficiently separated.
The slightly complicated anticommutation relations
(found for sufficiently close reference sites)
reflect the use of basis states which are
localized but not orthogonal with each other.

An important property of the \( a \) and
\( b \) operators is that one can represent
arbitrary states of the system by using
these operators.
The key is the following lemma.
\begin{lemma}
	For any \( r\in\Lambda \)
	and \( \sigma=\up,\dn \),
	one has
	\begin{equation}
		c_{r,\sigma}=\sum_{x\in\calE}\gamma_{x}\,\axs
		+\sum_{u\in\calI}\eta_{u}\,\bus,
		\label{eq:cab}
	\end{equation}
	with suitable coefficients 
	\( \gamma_{x} \) and \( \eta_{u} \).
\end{lemma}
\proof
Consider a Hilbert space which consists
of operators of the form
\( \sum_{r\in\Lambda}\alpha_{r}\,c_{r,\sigma} \)
with \( \alpha_{r}\in\comp \).
We fix \( \sigma \) to be
either \( \up \) or \( \dn \).
The inner product of the two ``vectors''
\( \sum_{r\in\Lambda}\alpha_{r}\,c_{r,\sigma} \)
and
\( \sum_{r\in\Lambda}\beta_{r}\,c_{r,\sigma} \)
is defined to be
the anticommutator
\( \{
(\sum_{r\in\Lambda}\alpha_{r}\,c_{r,\sigma})^\dagger,
\sum_{r\in\Lambda}\beta_{r}\,c_{r,\sigma}
\}
=
\sum_{r\in\Lambda}\overline{\alpha_{r}}\,\beta_{r}\).
Since
\( \{\adxs,\bus\}=0 \)
for any \( x\in\calE \)
and any \( u\in\calI \),
the subspace spanned by the
set \( \{\axs\}_{x\in\calE} \)
and that spanned by the
set \( \{\bus\}_{u\in\calI} \)
are orthogonal.
Since \( \axs \) with different \( x \) are
linearly independent, the dimension of
the former subspace is equal to \( |\calE| \).
Similarly the dimension of the latter subspace is 
\( |\calI| \).
Noting that \( |\calE|+|\calI|=|\Lambda| \)
is the dimension of the whole space,
one finds that the set
\( \{\axs\}_{x\in\calE}\cup\{\bus\}_{u\in\calI} \)
spans the whole space.
This means that any \( c_{r,\sigma} \)
can be expanded in terms of
\( \axs \) and \( \bus \)
as in (\ref{eq:cab}).\qed

Recall that the basis states of the many-electron
Hilbert space are (\ref{eq:basis}).
As a consequence of the lemma,
we find that an arbitrary
many-electron state of the system
can be represented as a linear combination of
the basis states
\begin{equation}
	\Psi_{0}^{(\nu)}(\Eup,\Edn,\Iup,\Idn)
	=
	\rbk{\prod_{x\in E_{\up}}a^\dagger_{x,\up}}
	\rbk{\prod_{x\in E_{\dn}}a^\dagger_{x,\dn}}
	\rbk{\prod_{u\in I_{\up}}b^\dagger_{u,\up}}
	\rbk{\prod_{u\in I_{\dn}}b^\dagger_{u,\dn}}
	\vac,
	\label{eq:Pexp}
\end{equation}
with arbitrary subsets 
\( E_{\up},E_{\dn}\subset\calE \)
and
\( I_{\up},I_{\dn}\subset\calI \).
Here
\( |E_{\up}|+|E_{\dn}|+|I_{\up}|+|I_{\dn}|=N_{\rm e} \)
is the total electron number.

\subsection{Definition of the model and the main theorem}
\label{s:defmain}
Our model is characterized by the four 
parameters 
\( t>0 \), \( s>0 \), \( U>0 \), and \( \nu>0 \).
The Hamiltonian of the model on \( \La \) is
\begin{equation}
	H
	=
	-s\sumtwo{x\in\calE}{\sigma=\up,\dn}
	a^\dagger_{x,\sigma}\,\axs
	+t\sumtwo{u\in\calI}{\sigma=\up,\dn}
	b^\dagger_{u,\sigma}\,\bus
	+U\sum_{r\in\La}n_{r,\up}n_{r,\dn},
	\label{eq:H}
\end{equation}
where the number operator \( n_{x,\sigma} \)
is defined in (\ref{eq:ndef}).

\begin{figure}
\centerline{\epsfig{file=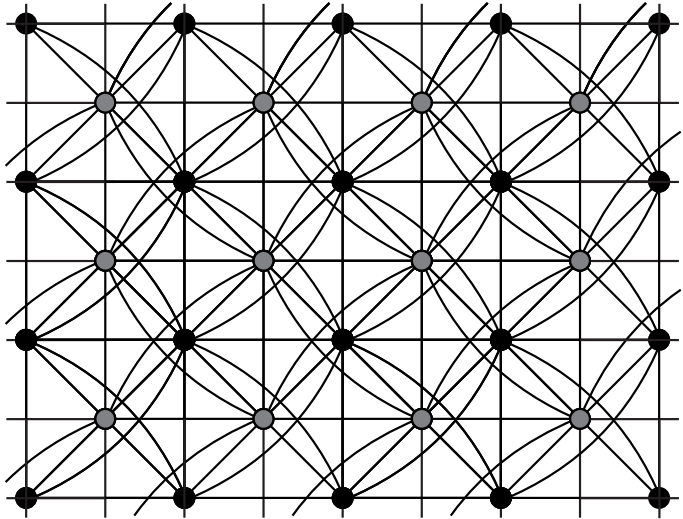,width=9cm}}
\caption[dummy]{
Another example on the fcc like lattice in
two dimensions obtained by
setting \( n=m=4 \).
}
\label{f:another}
\end{figure}

Recalling the definitions 
(\ref{eq:axs}) and (\ref{eq:bus}),
one sees that this defines a Hubbard model 
with nearest and next nearest neighbor
hopping terms.
We can rewrite (\ref{eq:H}) in the 
standard form (\ref{eq:genH}) 
with the hopping matrix given by
\begin{equation}
	\cases{
	t_{x,x}=mt\nu^2-s,&\( x\in\calE \)\cr
	t_{u,u}=t-ns\nu^2,&\( u\in\calI \)\cr
	t_{x,u}=t_{u,x}=\cases{
	\nu(t+s),&\( x\in C_{u} \)\cr
	0,&\( x\not\in C_{u} \)\cr
	}&\( x\in\calE \), \( u\in\calI \)\cr
	t_{x,y}=\ell_{x,y}\,\nu^2t,
	&\( x,y\in\calE \), \( x\ne y \)
	\cr
	t_{u,v}=-\ell_{u,v}\,\nu^2s,&
	\( u,v\in\calI \), \( u\ne v \)\cr
	}
	\label{eq:trs}
\end{equation}
Note that the model has nearest and next-nearest
neighbor hopping amplitudes, but not more.
See Figs.~\ref{f:1Dnf}, \ref{f:2Dlattice}~(b),
and \ref{f:another} for examples\footnote{
Observe that the lattice in Fig.~\ref{f:2Dlattice}~(b)
can be obtained by either setting \( n=2 \), \( m=4 \),
or \( n=4 \), \( m=2 \).
(In the latter case, the black dots correspond to
the internal sites.)
This means that we have models which exhibit saturated
ferromagnetism at different electron numbers
in different regions in the parameter space.
}.

We consider the Hilbert space with the 
electron number
fixed to \( N_{\rm e}=|\calE|=nM/m \).
Note that this electron number is
consistent with the interpretation that an
external site represents a metallic atom
which emits one electron to the system.

Exactly as in Theorem~\ref{t:flat},
it can be shown that the flat-band models
with \( s=0 \) exhibit saturated ferromagnetism
for any \( t>0 \), \( \nu>0 \) and \( U>0 \).
See Section~6 of \cite{PTP} for a proof.
The instability of saturated ferromagnetism
for sufficiently small \( U \) as in 
Theorem~\ref{t:non-f} can be
of course proved for the general models.
See Section~3.3 of \cite{PTP}.

Our main result is 
the following theorem which 
shows that the ground states of the model
exhibit saturated ferromagnetism\footnote{
Theorem~\ref{t:SW} about the low-lying excitation 
is valid in
a wide range of models.
See \cite{JSP}.
}.

\begin{theorem}
\label{t:main}
When \( t/s \), \( U/s \) and
\( 1/\nu \) are sufficiently large
(how large these quantities should be depends only
on the local structure of the lattice, but 
not on the size of the lattice),
the ground state of the model is 
\( (N_{\rm e}+1) \)-fold degenerate
and has the total spin 
\( \Stot=\Ne/2 \).
\end{theorem}	

In the proof of the theorem, we further show
that one of the ground states is
written as
\begin{equation}
	\Phi_{\rm GS}
	=
	\rbk{\prod_{x\in\calE}a^\dagger_{x,\up}}
	\vac,
	\label{eq:GS}
\end{equation}
and other ground states are obtained by 
applying the spin lowering operator
\( \hat{S}^-_{\rm tot}=
\sum_{r\in\La}c^\dagger_{r,\dn}c_{r,\up} \)
onto the state (\ref{eq:GS}).

\subsection{``Band'' structure in the single-electron
problem}
\label{s:band}
Before proceeding to prove the theorem, 
we discuss a basic property of the single electron
problem associated with the present model.
Although the discussion is not necessary
for the proof of the main theorem,
it sheds light on a special character of the
model that we are studying.

The single electron Hilbert space
\( \hilb \) is the
\( |\Lambda| \)-dimensional linear space spanned by
\( c^\dagger_{r,\up}\vac \) with \( r\in\Lambda \).
(We here consider the space of up-spin electrons,
but this choice is not essential.)
This space is decomposed as
\begin{equation}
	\hilb=\hilb_{\rm L}\oplus\hilb_{\rm U},
	\label{eq:hilbdec}
\end{equation}
where \( \hilb_{\rm L} \)
is spanned by \( a^\dagger_{x,\up}\vac \)
with \( x\in\calE \),
and \( \hilb_{\rm U} \)
by \( b^\dagger_{u,\up}\vac \)
with \( u\in\calI \).
Then we have the following.

\begin{pro}
	The Hamiltonian \( H \) can be diagonalized
	within \( \hilb_{\rm L} \) and
	within \( \hilb_{\rm U} \), respectively.
	The energy eigenvalues \( \epsilon \)
	in \( \hilb_{\rm L} \) satisfy
	\begin{equation}
		-s\{1+(m+\ell_{\rm L})\nu^2\}
		\le\epsilon\le
		\min\{0,
		-s\{1+(m-\ell_{\rm L})\nu^2\}\},
		\label{eq:epL}
	\end{equation}
	where 
	\( \ell_{\rm L}=\sum_{y\in\calE, y\not=x}\ell_{x,y} \)
	with \( x\in\calE \),
	and the energy eigenvalues \( \epsilon \)
	in \( \hilb_{\rm U} \) satisfy
	\begin{equation}
		\max\{0,
		t\{1+(n-\ell_{\rm U})\nu^2\}\}
		\le\epsilon\le
		t\{1+(n+\ell_{\rm U})\nu^2\},
		\label{eq:epH}
	\end{equation}
	where 
	\( \ell_{\rm U}=\sum_{v\in\calI, v\not=u}\ell_{u,v} \)
	with \( u\in\calI \).
\end{pro}

The proposition states that 
the spectrum of the Hamiltonian \( H \)
in the single electron Hilbert
space \( \hilb \) consists of two distinct
``bands.''
When \( \nu \) is sufficiently small (which is
the case when the theorem holds),
the two ``bands'' do not overlap
and are separated by a finite gap.
The fermion operator
\( a^\dagger_{x,\sigma} \)
creates an electron in the lower ``band'',
and \( b^\dagger_{u,\sigma} \)
creates an electron in the upper ``band.''

When the model has a translation invariance
as in the models of Section~\ref{s:typ},
the single electron Hilbert space is decomposed 
into several bands in the standard sense.
The lower or upper ``band'' that we mentioned above
is not necessarily a band in the usual sense, but
may be a union of several bands.
In the band structure (\ref{eq:ejk}) discussed in
Section~\ref{s:typ}, the lowest band with 
\( j=1 \) corresponds to the lower ``band'',
and the collection of the remaining 
\( d \) bands corresponds to the upper ``band.''

\proof
The proof is elementary but requires some care.

Consider a state \( \Phi \)
of the form
\begin{equation}
	\Phi=\sum_{x\in\calE}
	\varphi(x)a^\dagger_{x,\up}\vac,
	\label{eq:Phis}
\end{equation}
where \( \varphi(x) \) are complex coefficients.
From the anticommutation relations
(\ref{eq:aa}), one finds that
\begin{equation}
	H\Phi=-\sum_{x\in\calE}s
	\cbk{
	(1+m\nu^2)\varphi(x)
	+\nu^2\sumtwo{y\in\calE}{y\not=x}
	\ell_{x,y}\varphi(y)
	}a^\dagger_{x,\up}\vac.
	\label{eq:HPhis}
\end{equation}
Since the right-hand side is again a linear
combination of 
\( a^\dagger_{x,\up}\vac \),
we find that \( H \) can be diagonalized
within \( \hilb_{\rm L} \).
We now assume \( H\Phi=\epsilon\Phi \).
By comparing the coefficients in
(\ref{eq:Phis}) and (\ref{eq:HPhis}),
we find
\begin{equation}
	-s(1+m\nu^2)\varphi(x)
	-s\nu^2\sumtwo{y\in\calE}{y\not=x}
	\ell_{x,y}\varphi(y)
	=
	\epsilon\,\varphi(x).
	\label{eq:Sch}
\end{equation}
By multiplying (\ref{eq:Sch}) with
\( \overline{\varphi(x)} \),
by summing it over \( x\in\calE \),
and by solving it for \( \epsilon \),
one gets
\begin{equation}
	\epsilon
	=
	-\rbk{\sum_{x\in\calE}|\varphi(x)|^2}^{-1}
	\rbk{
	s(1+m\nu^2)\sum_{x\in\calE}|\varphi(x)|^2
	+s\nu^2\sumtwo{x,y\in\calE}{x\not=y}
	\ell_{x,y}\,\overline{\varphi(x)}\,\varphi(y)
	}.
	\label{eq:epeq}
\end{equation}
By using the inequalities
\begin{equation}
	-(|\varphi(x)|^2+|\varphi(y)|^2)
	\le
 	\overline{\varphi(x)}\,\varphi(y)+\varphi(x)\,\overline{\varphi(y)}
	\le
	|\varphi(x)|^2+|\varphi(y)|^2,
	\label{eq:easy}
\end{equation}
which follow immediately from
\( |\varphi(x)\pm\varphi(y)|^2\ge0 \),
we find from (\ref{eq:epeq}) that
\begin{equation}
	-s(1+m\nu^2)-s\nu^2\ell_{\rm L}
	\le
	\epsilon
	\le-s(1+m\nu^2)+s\nu^2\ell_{\rm L}.
	\label{eq:eple}
\end{equation}
This, with the positive semidefiniteness of
\( a^\dagger_{x,\sigma}a_{x,\sigma} \),
proves the desired (\ref{eq:epL}).
The other inequality
(\ref{eq:epH}) is proved in exactly the same manner
using the \( b^\dagger_{u,\up} \)
operators.\qed

\section{Proof}
\label{s:proof}

\subsection{Proof of the main theorem}
\label{s:pr}
The basic strategy of the proof is first to
show the appearance of ferromagnetism in a local piece
of the system, and then to ``connect'' these local
ferromagnetism together to get the
desired ferromagnetic state on the whole system.
Of course such a ``connection'' usually does not work in 
itinerant electron systems where electrons behave
as ``waves.''
Our method makes a full use of special features of our
model as well as of ferromagnetic states.

Our proof is based on the decomposition of
the Hamiltonian
\begin{equation}
	H=\sum_{x\in\calE}h_{x},
	\label{eq:Hh}
\end{equation}
where \( h_{x} \) acts only on the sublattice
\( \La_{x} \).
The local Hamiltonian 
\( h_{x} \) is defined as
\begin{equation}
	h_{x}=
	-s\sum_{\sigma=\up,\dn}
	a^\dagger_{x,\sigma}\axs
	+\frac{t}{n}
	\sumtwo{u\in\La_{x}\cap\calI}{\sigma=\up,\dn}
	b^\dagger_{u,\sigma}\bus
	+\frac{U}{n'}
	\sum_{y\in\La_{x}\cap\calE}
	n_{y,\up}n_{y,\dn}
	+\frac{U}{n}
	\sum_{u\in\La_{x}\cap\calI}
	n_{u,\up}n_{u,\dn},
	\label{eq:hx}
\end{equation}
where \( n'=|\La_{x}\cap\calE| \).
It should be stressed that 
\( h_{x} \) with neighboring \( x \)
do not commute with each other.
One therefore cannot diagonalize all 
\( h_{x} \) simultaneously.

As for the lowest eigenvalue
and the corresponding eigenstates, however,
we have the following.
This lemma plays a key role in our proof
of the theorem.

\begin{lemma}
\label{lemma}
When \( t/s \), \( U/s \) and
\( 1/\nu \) are sufficiently large,
the lowest eigenvalue of \( h_{x} \)
is \( -s(1+m\nu^2) \), and any corresponding
eigenstate \( \Phi \) can always be written as
\begin{equation}
	\Phi=a^\dagger_{x,\up}\Phi_{1}
	+a^\dagger_{x,\dn}\Phi_{2},
	\label{eq:PaPaP}
\end{equation}
where \( \Phi_{1} \), \( \Phi_{2} \)
are suitable states with \( N_{\rm e}-1 \)
electrons.
The eigenstate \( \Phi \) also
satisfies
\begin{equation}
	c_{r,\up}c_{r,\dn}\Phi=0,
	\label{eq:ccP1}
\end{equation}
for any \( r\in\La_{x} \).
\end{lemma}

We shall prove Theorem~\ref{t:main} 
assuming Lemma~\ref{lemma}.
The lemma will be proved
in Section~\ref{s:prLemma}.

Since \( h_{x}\ge-s(1+m\nu^2) \),
we have 
\( H=\sum_{x\in\calE}h_{x}
\ge-s(1+m\nu^2)|\calE| \).
A straightforward calculation 
using the anticommutation relations
(\ref{eq:ab}) and (\ref{eq:aa}) shows that
the state (\ref{eq:GS}) is an eigenstate
of \( H \) with the eigenvalue
\( -s(1+m\nu^2)|\calE| \).
Therefore we see that the state (\ref{eq:GS}) 
is a ground state.
Our goal here is to characterize
all the ground states.

Let \( \Phi \) be an arbitrary eigenstate
of \( H \) with the eigenvalue
\( -s(1+m\nu^2)|\calE| \).
Then it follows from  
\( h_{x}\ge-s(1+m\nu^2) \)
that
\begin{equation}
	h_{x}\Phi=-s(1+m\nu^2)\Phi,
	\label{eq:hxP}
\end{equation}
for any \( x\in\calE \).
Thus \( \Phi \) satisfies the properties
stated in Lemma~\ref{lemma}.

Let us expand \( \Phi \)
in the basis states
\( \Psi_{0}^{(\nu)}(\Eup,\Edn,\Iup,\Idn) \) of
(\ref{eq:Pexp}).
Since \( \Phi \) satisfies
(\ref{eq:PaPaP}) for any \( x\in\calE \),
it follows that only those 
basis states
with
\( E_{\up}\cup E_{\dn}=\calE \)
contribute.
Since the electron number is
\( |\Eup|+|\Edn|+|\Iup|+|\Idn|=N_{\rm e}
=|\calE| \),
the condition
\( E_{\up}\cup E_{\dn}=\calE \)
implies 
\( E_{\up}\cap E_{\dn}=\emptyset \)
and
\( I_{\up}=I_{\dn}=\emptyset \).
Therefore the expansion
of \( \Phi \) in the basis states (\ref{eq:Pexp})
can be rearranged into a ``spin system
representation'' as
\begin{equation}
	\Phi=\sum_{\vsigma}\psi(\vsigma)
	\rbk{\prod_{x\in\calE}
	a^\dagger_{x,\sigma(x)}}
	\vac,
	\label{eq:Ppsi}
\end{equation}
where \( \vsigma=(\sigma(x))_{x\in\calE} \)
is summed over all the spin configurations
with \( \sigma(x)=\up,\dn \),
and  \( \psi(\vsigma) \)
are complex coefficients.

We then examine the property (\ref{eq:ccP1})
for \( u\in\calI \).
From the definition
(\ref{eq:axs}), we find that
for any \( u\in\calI \)
\begin{equation}
	c_{u,\up}c_{u,\dn}
	\rbk{\prod_{x\in\calE}
	a^\dagger_{x,\sigma(x)}}
	\vac
	=
	\sumtwo{y,z\in C_{u}\exfor{u}}{y\not=z}
	{\rm sgn}(y,z)
	\,
	\chi[\sigma(y)=\up,\sigma(z)=\dn]
	\rbk{\prod_{x\in\calE\exfor{y,z}}
	a^\dagger_{x,\sigma(x)}}
	\vac,
	\label{eq:ccaP}
\end{equation}
where the sign factor \( {\rm sgn}(y,z) \)
comes from the anticommutation relations,
and satisfies 
\( {\rm sgn}(y,z)=-{\rm sgn}(z,y) \).
The characteristic function \( \chi[\cdot] \)
is defined as usual by
\( \chi[{\rm true}]=1 \)
and
\( \chi[{\rm false}]=0 \).
By using (\ref{eq:Ppsi}) and (\ref{eq:ccaP}), we 
find  for any \( u\in\calI \) that
\begin{equation}
	c_{u,\up}c_{u,\dn}\Phi
	=
	\sumtwo{y,z\in C_{u}\exfor{u}}
	{y>z}
	\sum_{\vsigma}
	{\rm sgn}(y,z)
	\,
	\chi[\sigma(y)=\up,\sigma(z)=\dn]
	\,
	\{\psi(\vsigma)-
	\psi(\vsigma_{y\rightleftharpoons z})\}
	\rbk{\prod_{x\in\calE\exfor{y,z}}
	a^\dagger_{x,\sigma(x)}}
	\vac,
	\label{eq:ccPP}
\end{equation}
where we have introduced an arbitrary ordering in
\( \calE \) to avoid double counting.
The spin configuration 
\( \vsigma_{y\rightleftharpoons z} \)
is obtained from 
\( \vsigma=(\sigma(x))_{x\in\calE} \)
by switching 
\( \sigma(y) \) and \( \sigma(z) \).
Since the basis states in the sum
(\ref{eq:ccPP}) are all linearly independent,
we find from the property (\ref{eq:ccP1}) that
\begin{equation}
	\psi(\vsigma)
	=
	\psi(\vsigma_{y\rightleftharpoons z}),
	\label{eq:psps}
\end{equation}
for any \( y,z\in\calE \) for which 
there is \( u\in\calI \) such that
\( y,z\in C_{u} \).
Since the whole lattice is connected,
(\ref{eq:psps}) implies that all 
\( \psi(\vsigma) \)
with the same 
\( M=\sum_{x\in\calE}\sigma(x) \)
are identical.
This completes the characterization of the 
space of the ground states.
The ground state which has a fixed
total spin in the \( z \)-direction
is
\begin{equation}
	\Phi_{M}=
	\sumtwo{\vsigma}{\sum_{x}\sigma(x)=M}
	\rbk{\prod_{x\in\calE}
	a^\dagger_{x,\sigma(x)}}
	\vac,
	\label{eq:PM}
\end{equation}
where 
\( M=-(|\calE|/2), 1-(|\calE|/2), 
,\cdots, (|\calE|/2)-1, |\calE|/2\).
Thus the ground states are \( |\calE|+1 \) fold degenerate.
It is easy to check that 
\begin{equation}
	(\hat{\bf S}_{\rm tot})^2
	\Phi_{M}
	=\Smax(\Smax+1)\Phi_{M},
	\quad
	\hat{S}^{(z)}_{\rm tot}\Phi_{M}
	=
	M\,\Phi_{M},
	\label{eq:SPM}
\end{equation}
with \( \Smax=|\calE|/2 \) being the maximum possible
value of the total spin of \( \Ne=|\calE| \) electrons.

\subsection{Some extensions}
\label{s:ext}
Let us make two brief remarks about 
extensions of Theorem~\ref{t:main}.

The first extension deals with the one dimensional
model of Fig.~\ref{f:1Dnf}, which 
(in the notation of Section~\ref{s:lattice})
has \( n=m=2 \).
In this model,
Tanaka \cite{TanakaPC} observed that 
the statement of Theorem~\ref{t:main}
can be proved if one first fixes arbitrary
\( \nu>0 \) and then takes sufficiently
large \( t/s \) and \( U/s \).

To show this extended theorem, one proves
the statement corresponding to
Lemma~\ref{lemma} by the method we used
in \cite{nonflat} to prove the similar
lemma.
The differences between the lemma in 
\cite{nonflat} and that in the present paper
comes from a difference in the definitions
of the local Hamiltonian.
Unlike the definition (\ref{eq:hx}) in the present paper, 
we did not include  the on-site repulsion terms from the 
external sites other than \( x \) in
the local Hamiltonian used in \cite{nonflat}.
This seemingly minor difference indeed makes a considerable
difference in the conditions that we obtain
in the limit \( U\to\infty \).
The same method as in \cite{nonflat} thus yields
much stronger information for the local Hamiltonian
defined as in the present paper\footnote{
After the publication of \cite{nonflat},
Kubo \cite{KuboPC} and Shen \cite{ShenPC,Shen98} 
independently
noticed the importance to include the on-site 
repulsions from the external sites in the local
Hamiltonian.
}.
We leave the details to 
the interested readers.

The second extension is much more straightforward
and less important.
For arbitrary complex coefficients 
\( f_{u} \), define
\begin{equation}
	B=\sum_{\sigma=\up,\dn}
	\cbk{\sum_{u\in\calI}f_{u}b_{u,\sigma}}^\dagger
	\cbk{\sum_{u\in\calI}f_{u}b_{u,\sigma}},
	\label{eq:B}
\end{equation}
which is obviously positive semidefinite.
From the expression (\ref{eq:SPM}) for the ground
states and the anticommutation relations
(\ref{eq:ab}),
one readily finds that \( B\Phi=0 \)
for any ground state \( \Phi \).

This means that one may add to the Hamiltonian
the new hopping terms
\begin{equation}
	H'=\sum_{j}\sum_{\sigma=\up,\dn}
	\cbk{\sum_{u\in\calI}f_{u}^{(j)}b_{u,\sigma}}^\dagger
	\cbk{\sum_{u\in\calI}f_{u}^{(j)}b_{u,\sigma}},
	\label{eq:H'}
\end{equation}
with arbitrary \( f^{(j)}_{u} \)
without modifying the ferromagnetic ground states.
In this manner, one can modify, for example, the models
in Section~\ref{s:typ} so that all the bands become dispersive
maintaining the appearance (and the provability)
of saturated ferromagnetism.

\subsection{Proof of Lemma~\protect\ref{lemma}}
\label{s:prLemma}
It suffices to prove the lemma for \( h_{o} \)
with a fixed \( o\in\calE \).
Since \( h_{o} \) acts only on
\( \Lao \), we only consider an electron
system defined on \( \Lao \) without
specifying the electron number.
We also write
\( \Eo=\Lao\cap\calE \)
and 
\( \Io=\Lao\cap\calI \).
The local Hamiltonian that we consider is
\begin{equation}
	h_{o}=
	-s\sum_{\sigma=\up,\dn}
	a^\dagger_{o,\sigma}a_{o,\sigma}
	+\frac{t}{n}
	\sumtwo{u\in\Io}{\sigma=\up,\dn}
	b^\dagger_{u,\sigma}\bus
	+\frac{U}{n'}
	\sum_{x\in\Eo}
	n_{x,\up}n_{x,\dn}
	+\frac{U}{n}
	\sum_{u\in\Io}
	n_{u,\up}n_{u,\dn}.
	\label{eq:ho}
\end{equation}

We stress that the statement of the Lemma is
about the property of a finite dimensional
matrix \( h_{o} \).
It is thus possible (in principle) to prove the lemma
for fixed \( n \), \( m \) by using a computer\footnote{
The numerical values of \( t/s \) in the caption to
Fig.~\ref{f:1Dnf} was obtained by using a computer.
See also Fig.~2 of \cite{nonflat}.
Shen \cite{Shen98} has done this for some models in higher 
dimensions.
}.
But the problem is indeed rather delicate,
and the proof for general cases seems
highly nontrivial.

As we have restricted our lattice,
we redefine (only in this proof)
the operator \( a_{x,\sigma} \)
for \( x\in\Eo\exfor{o} \) as
\begin{equation}
	a_{x,\sigma}
	=
	c_{x,\sigma}-
	\nu\sum_{u\in\La_{x}\cap\Io}
	c_{u,\sigma}.
	\label{eq:ays2}
\end{equation}
The definition of \( b_{u,\sigma} \)
is unchanged.
Note that we still have 
\( \{a^\dagger_{x,\sigma},b_{u,\tau}\}=0 \).

Exactly as in (\ref{eq:Pexp}),
any state defined on \( \Lao \)
can be written as a linear combination
of the basis states
\begin{equation}
	\Psi_{1}^{(\nu)}(\Eup,\Edn,\Iup,\Idn)
	=
	\rbk{\prod_{x\in\Eup}a^\dagger_{x,\up}}
	\rbk{\prod_{x\in\Edn}a^\dagger_{x,\dn}}
	\rbk{\prod_{u\in\Iup}b^\dagger_{u,\up}}
	\rbk{\prod_{u\in\Idn}b^\dagger_{u,\dn}}
	\vac,
	\label{eq:Psi1}
\end{equation}
with arbitrary subsets
\( \Eup,\Edn\subset\Eo \)
and
\( \Iup,\Idn\subset\Io \).
Here we do not fix the electron number
which is given by
\( |\Eup|+|\Edn|+|\Iup|+|\Idn| \).

\subsubsection{The limit $t,U\to\infty$}
\label{sec:tU}
Let us first consider the limit where
\( t\to\infty \) and \( U\to\infty \).
It is easily found that
the lowest 
eigenvalue 
of \( h_{o} \) is finite
in this limit.
(Try, for example, the state
\( a^\dagger_{o,\up}\vac \).)
Note that the parts in 
\( h_{o} \) which contain \( t \)
are
\( (t/n)
	\sum_{u\in\Io}\sum_{\sigma=\up,\dn}
	b^\dagger_{u,\sigma}\bus \),
and which contain \( U \) are
\( (U/n')
	\sum_{x\in\Eo}
	n_{x,\up}n_{x,\dn}
	+(U/n)
	\sum_{u\in\Io}
	n_{u,\up}n_{u,\dn} \).
Since each term in these sums is
positive semidefinite,
the necessary and sufficient condition for
a state \( \Phi \) to have a finite energy
in the limit \( t\to\infty \), \( U\to\infty \)
is
\begin{equation}
	b_{u,\sigma}\Phi=0,
	\label{eq:bP}
\end{equation}
for any \( u\in\Io \) 
and \( \sigma=\up,\dn \),
and
\begin{equation}
	c_{r,\up}c_{r,\dn}\Phi=0
	\label{eq:ccP}
\end{equation}
for any \( r\in\Lao \).
To get (\ref{eq:ccP}),
we noted that
\( n_{r,\up}n_{r,\dn}=
(c_{r,\up}c_{r,\dn})^\dagger 
c_{r,\up}c_{r,\dn}\).

To see implications of 
the condition (\ref{eq:bP}),
we introduce dual operators
\( \tilde{b}_{u,\sigma} \)
for \( u\in\Io \) and
\( \sigma=\up,\dn \)
which satisfy
\begin{equation}
	\{\tilde{b}_{u,\sigma},b^\dagger_{v,\tau}\}
	=
	\delta_{u,v}\,\delta_{\sigma,\tau},
	\label{eq:tbb}
\end{equation}
for any \( u,v\in\Io \)
and \( \sigma,\tau=\up,\dn \).
More precisely the construction is as follows.
Define the Gramm matrix \( \mG \)
by 
\( (\mG)_{u,v}=
\{b_{u,\sigma},b^\dagger_{v,\sigma}\} \),
where \( u,v \) run over 
\( \Io \).
The linear independence of the basis states
implies that \( \mG \) is invertible.
For \( u\in\Io \)
and \( \sigma=\up,\dn \), define
\begin{equation}
	\tilde{b}_{u,\sigma}
	=
	\sum_{w\in\Io}
	(\mG^{-1})_{u,w}\,b_{w,\sigma},
	\label{eq:tbus}
\end{equation}
where it is easy to check (\ref{eq:tbb}).

From (\ref{eq:tbb}) and (\ref{eq:Psi1}),
one sees that
\begin{equation}
	b^\dagger_{u,\sigma}\,
	\tilde{b}_{u,\sigma}
	\Psi_{1}^{(\nu)}(\Eup,\Edn,\Iup,\Idn)
	=
	\cases{
	\Psi_{1}^{(\nu)}(\Eup,\Edn,\Iup,\Idn),&
	if \( u\in I_{\sigma} \);\cr
	0,&otherwise,
	}
	\label{eq:bbP1}
\end{equation}
for any \( u\in\Io \) and \( \sigma=\up,\dn \).
Let \( \Phi \) be an arbitrary state satisfying
(\ref{eq:bP}).
Then since \( \tilde{b}_{u,\sigma} \)
is a linear combination of
\( b_{w,\sigma} \),
one has
\( \tilde{b}_{u,\sigma}\Phi=0 \)
and hence
\begin{equation}
	b^\dagger_{u,\sigma}\tilde{b}_{u,\sigma}\Phi=0,
	\label{eq:bbP0}
\end{equation}
for any \( u\in\Io \)
and \( \sigma=\up,\dn \).
Then from (\ref{eq:bbP1}) and the linear
independence of the basis states
(\ref{eq:Psi1}),
one finds that the state \( \Phi \),
when expanded in the basis states
\( \Psi_{1}^{(\nu)} \),
cannot include
\( \Psi_{1}^{(\nu)}(\Eup,\Edn,\Iup,\Idn) \)
with nonempty
\( \Iup \) or \( \Idn \).

Therefore we conclude that \( \Phi \)
is a linear combination of the basis states
\begin{equation}
	\Psi_{2}^{(\nu)}(\Eup,\Edn)
	=
	\Psi_{1}^{(\nu)}(\Eup,\Edn,\emptyset,\emptyset)
	=
	\rbk{\prod_{x\in\Eup}a^\dagger_{x,\up}}
	\rbk{\prod_{x\in\Edn}a^\dagger_{x,\dn}}
	\vac,
	\label{eq:Psi2}
\end{equation}
with arbitrary \( \Eup,\Edn\subset\Eo \).

We then examine the condition
(\ref{eq:ccP}) for \( r\in\Eo \).
Noting the definitions 
(\ref{eq:axs}), (\ref{eq:ays2})
of \( \axs \) and (\ref{eq:Psi2}), we see that
\begin{equation}
	a^\dagger_{x,\dn}a^\dagger_{x,\up}
	c_{x,\up}c_{x,\dn}
	\Psi_{2}^{(\nu)}(\Eup,\Edn)
	=
	\cases{
	\Psi_{2}^{(\nu)}(\Eup,\Edn),
	&if \( x\in\Eup\cap\Edn \);\cr
	0,&otherwise,
	}
	\label{eq:aaccP2}
\end{equation}
for any \( x\in\Eo \).
Now the condition 
(\ref{eq:ccP}) for \( r\in\Eo \)
implies
\begin{equation}
	a^\dagger_{x,\dn}a^\dagger_{x,\up}
	c_{x,\up}c_{x,\dn}\Phi=0,
	\label{eq:aaccPhi}
\end{equation}
for any
\( x\in\Eo \).
Then, as before, we see that
\( \Phi \) is a linear combination of 
\( 	\Psi_{2}^{(\nu)}(\Eup,\Edn) \) 
with \( \Eup,\Edn\subset\Eo \) such that
\( \Eup\cap\Edn=\emptyset \).

For the state \( \Phi \) to have a finite
energy in the limits 
\( t,U\to\infty \),
it must further satisfy
\begin{equation}
	c_{u,\up}c_{u,\dn}\Phi=0
	\label{eq:zec}
\end{equation}
for any \( u\in\Io \).
This condition is not as straightforward to
treat as the previous two conditions.

To see implications of (\ref{eq:zec}),
we first note that when \( \Eup\cap\Edn=\emptyset \),
\begin{equation}
	c_{u,\up}c_{u,\dn}
	\Psi_{2}^{(\nu)}(\Eup,\Edn)
	=
	\nu^2
	\!\!
	\sumtwo{x,y\in C_{u}\exfor{u}}{x\not=y}
	\!\!
	\chi[x\in\Eup,y\in\Edn]
	\,
	{\rm sgn}(x,y;\Eup,\Edn)
	\,
	\Psi_{2}^{(\nu)}(\Eup\exfor{x},\Edn\exfor{y}),
	\label{eq:ccP2}
\end{equation}
where we used the 
definitions (\ref{eq:axs}), (\ref{eq:ays2})
of \( \axs \),
and (\ref{eq:Psi2}) of \( \Psi_{2}^{(\nu)}(\Eup,\Edn) \).
Here \( \chi[\cdot] \) is the characteristic
function as before, and
\( {\rm sgn}(x,y;\Eup,\Edn)=\pm1 \) is the
sign factor coming from anticommutation relations.

Let us then expand the state \( \Phi \) as
\begin{equation}
	\Phi=\sumtwo{\Eup,\Edn\subset\Eo}
	{\Eup\cap\Edn=\emptyset}
	\varphi(\Eup,\Edn)\,
	\Psi_{2}^{(\nu)}(\Eup,\Edn).
	\label{eq:Phiexp2}
\end{equation}
The zero energy condition 
(\ref{eq:zec}) for any \( u\in\Io \)
implies certain relations that the
coefficients \( \varphi(\Eup,\Edn) \)
must satisfy.
Noting that the parameter \( \nu \) appears in
(\ref{eq:ccP2}) only as a prefactor,
one finds that these relations for
\( \varphi(\Eup,\Edn) \) depend only on the lattice
structure and do not depend on \( \nu \)
at all.
Although the precise forms of the
relations are not needed here, let us write
them down for completeness.
The conditions that the 
coefficients \( \varphi(\Eup,\Edn) \)
must satisfy are
\begin{equation}
	\sumtwo{x,y\in C_{u}\exfor{u}}{x\not=y}
	\!\!
	\chi[x\not\in E'_{\up},y\not\in E'_{\dn}]
	\,
	{\rm sgn}(x,y;E'_{\up}\cup\{x\},E'_{\dn}\cup\{y\})
	\,
	\varphi(E'_{\up}\cup\{x\},E'_{\dn}\cup\{y\})
	=0,
	\label{eq:zecp}
\end{equation}
for any 
\( E'_{\up},E'_{\dn}\subset\Eo \)
such that 
\( E'_{\up}\cap E'_{\dn}=\emptyset \),
and for any \( u\in\Io \).

For \( \nu\ge0 \), we let
 \( \Hfn \) be the space of all \( \Phi \)
which are expanded as (\ref{eq:Phiexp2})
with the coefficients 
\( \varphi(\Eup,\Edn) \) satisfying the
conditions (\ref{eq:zecp}).
For \( \nu>0 \), the space
\( \Hfn \) is precisely the space of
all \( \Phi \) which have finite
energy (expectation value) in the 
limit \( t,U\to\infty \).
The space \( \Hfz \) has no such interpretation,
but it is convenient to define this space.
Note that \( \Hfn \) depends continuously
on \( \nu\ge0 \) since the range of 
allowed coefficients \( \varphi(\Eup,\Edn) \)
is independent of \( \nu \) and
the basis states \( \Psi_{2}^{(\nu)}(\Eup,\Edn) \)
are continuous in \( \nu \).

We also let \( \Pfn \) be the orthogonal
projection onto the space
\( \Hfn \).
Again \( \Pfn \) is continuous in
\( \nu\ge0 \).

For \( \nu>0 \), 
to study finite energy states of
the local Hamiltonian
\( h_{o} \) in the limit
\( t,U\to\infty \)
is equivalent to study
the effective Hamiltonian
\begin{equation}
	\hen=\Pfn\,\htn\,\Pfn,
	\label{eq:heff}
\end{equation}
where
\begin{equation}
	\htn=-s\sum_{\sigma=\up,\dn}
	a^\dagger_{o,\sigma}a_{o,\sigma}.
	\label{eq:tildeh}
\end{equation}
Again we extend the range of
\( \nu \) and define \( \hez \)
by (\ref{eq:heff}) with \( \nu=0 \).
Since \( \htn \) is continuous in
\( \nu\ge0 \), the effective Hamiltonian
\( \hen \) is also continuous in \( \nu\ge0 \).

It follows from the standard argument that
the eigenvalues of the local Hamiltonian
\( h_{o} \) with given \( \nu>0 \)
are classified into two sets.
In the limit \( t,U\to\infty \),
the eigenvalues in the first set
diverge, while
those in the second set converge to the eigenvalues
of \( \hen \) including the degeneracies.

Our next task is to investigate the 
eigenvalues of \( \hen \).
But this is still a nontrivial problem
since \( \Pfn \)
and \( \htn \)
do not commute.

\subsubsection{The case $\nu=0$}
Let us set \( \nu=0 \) and study \( \hez \).
Although \( \hez \) is not really an effective
Hamiltonian, we get crucial information about
\( \hen \) by studying 
\( \hez \).

For \( \nu=0 \), the operator
\( \axs \) is nothing but the basic fermion operator
\( \cxs \),
and the part (\ref{eq:tildeh}) of the local 
Hamiltonian becomes
\begin{equation}
	\htz=-s\,n_{o},
	\label{eq:nulim}
\end{equation}
where \( n_{o}=n_{o,\up}+n_{o,\dn} \)
is the number operator.
The problem becomes that of  electrons
strictly localized at sites in \( \Eo \),
except for the 
projection operator \( \Pfz \).
The existence of the projection makes
the problem nontrivial.

Let us decompose the space
\( \Hfz \) as
\begin{equation}
	\Hfz=\Gz\oplus\Vz\oplus\Mz.
	\label{eq:Hdecomp}
\end{equation}
Here \( \Gz \) consists of all 
\( \Phi\in\Hfz \) which 
satisfy \( n_{o}\Phi=\Phi \).
In other words,
\( \Gz \) is a set of states in 
\( \Hfz \) with singly occupied \( o \).
Any \( \Phi\in\Gz \)
is written as a 
linear combination of 
\( \Psi_{2}^{(0)}(\Eup,\Edn) \)
with \( \Eup\cup\Edn\ni o \).
Similarly \( \Vz \) consists of all
\( \Phi\in\Hfz \) which satisfy
\( n_{o}\Phi=0 \).
It is a set of states in 
\( \Hfz \) with vacant \( o \).
Any \( \Phi\in\Vz \) is written as a 
linear combination of 
\( \Psi_{2}^{(0)}(\Eup,\Edn) \)
with \( \Eup\cup\Edn\not\ni o \).
The space \( \Mz \) is defined as the 
orthogonal complement.

Note that \( \Gz \) and \( \Vz \)
are never empty since
\( c^\dagger_{o,\sigma}\vac\in\Gz \)
and
\( c^\dagger_{x,\sigma}\vac\in\Vz \)
where \( x\in\Eo\exfor{o} \).
On the other hand,
\( \Mz \) is empty in models with \( n=2 \).
Since the following argument becomes almost trivial
when \( \Mz  \) is empty, we shall assume that
\( \Mz \) is not empty.

Any \( \Phi\in\Mz \)
is uniquely decomposed as
\begin{equation}
	\Phi=
	c^\dagger_{o,\up}\Phi_{1}+
	c^\dagger_{o,\dn}\Phi_{2}+
	\Phi_{3},
	\label{eq:Phidec}
\end{equation}
where 
\( \Phi_{i} \) (\( i=1,2,3 \))
satisfy \( n_{o}\Phi_{i}=0 \),
i.e., 
\( o \) is vacant in these states.
We then define
\begin{equation}
	\alpha=
	\sup_{\Phi\in\Mz}
	\frac{\norm{c^\dagger_{o,\up}\Phi_{1}+
	c^\dagger_{o,\dn}\Phi_{2}}}
	{\norm{c^\dagger_{o,\up}\Phi_{1}+
	c^\dagger_{o,\dn}\Phi_{2}+
	\Phi_{3}}},
	\label{eq:alpha}
\end{equation}
and note that 
\begin{equation}
	\alpha<1.
	\label{eq:a<1}
\end{equation}
To see this, suppose that \( \alpha=1 \).
Since \( \Mz \) is closed there is 
\( \Phi\in\Mz \) which attains \( \alpha=1 \).
Then \( \alpha=1 \)
implies
\( \norm{c^\dagger_{o,\up}\Phi_{1}+
	c^\dagger_{o,\dn}\Phi_{2}+
	\Phi_{3}}
	=
	\norm{c^\dagger_{o,\up}\Phi_{1}+
	c^\dagger_{o,\dn}\Phi_{2}}\),
which means \( \Phi_{3}=0 \).
But this means \( \Phi\in\Gz \),
which contradicts with (\ref{eq:Hdecomp}).

Now from (\ref{eq:nulim}), one has
\begin{equation}
	\hez\Phi=\Pfz\htz\Pfz\Phi
	=-s\,\Pfz n_{o}\Phi
	=
	\cases{
	-s\,\Phi,&if \( \Phi\in\Gz \);\cr
	0,&if \( \Phi\in\Vz \).
	}
	\label{eq:heP}
\end{equation}
Thus \( \Phi \) in \( \Gz \) or \( \Vz \)
is an eigenstate of \( \hez \).
The remaining eigenstates are within the 
space \( \Mz \).

As for \( \Phi\in\Mz \),
one has
\begin{eqnarray}
	\hez\Phi&=&
	\Pfz\htz\Pfz\Phi
	\ret
	&=&
	-s\,\Pfz n_{o}
	(c^\dagger_{o,\up}\Phi_{1}+
	c^\dagger_{o,\dn}\Phi_{2}+
	\Phi_{3})
	\ret
	&=&
	-s\,\Pfz
	(c^\dagger_{o,\up}\Phi_{1}+
	c^\dagger_{o,\dn}\Phi_{2}).
	\label{eq:Mz}
\end{eqnarray}
Therefore
\begin{eqnarray}
	(\Phi,\hez\Phi)
	&=&
	-s(\Phi,\Pfz
	(c^\dagger_{o,\up}\Phi_{1}+
	c^\dagger_{o,\dn}\Phi_{2}))
	\ret
	&\ge&
	-s\norm{\Phi}\norm{c^\dagger_{o,\up}\Phi_{1}+
	c^\dagger_{o,\dn}\Phi_{2}},
	\label{eq:PhP}
\end{eqnarray}
and we get
\begin{equation}
	\frac{(\Phi,\hez\Phi)}{\norm{\Phi}^2}
	\ge
	-s\frac{\norm{c^\dagger_{o,\up}\Phi_{1}+
	c^\dagger_{o,\dn}\Phi_{2}}}{\norm{\Phi}}
	\ge
	-\alpha s
	>
	-s,
	\label{eq:enlob}
\end{equation}
where we used (\ref{eq:alpha}) and (\ref{eq:a<1}).
By the variational principle, we see that
the eigenvalues of \( \hez \) within
the space \( \Mz \) are not less than
\( -\alpha s \).

Thus we found that the lowest eigenvalue of
\( \hez \) is \( -s \) and its degeneracy is
equal to the dimension \( \gamma \) 
of the space \( \Gz \).
Note that \( \gamma \) is not vanishing
since \( \Gz \) is not empty.
There is a finite gap above the
lowest eigenvalue.

\subsubsection{Non-limiting cases}
By using the properties of \( \hez \) and 
the continuity in \( \nu\ge0 \),
one finds that \( \hen \) has 
\( \gamma \) low lying eigenvalues 
which are separated from larger eigenvalues
by a finite gap, provided that \( \nu>0 \)
is sufficiently small.
By recalling the remark at
the end of section~\ref{sec:tU}, 
one finds that
for sufficiently small 
\( \nu>0 \) and sufficiently large
\( t \) and \( U \),
the local Hamiltonian 
\( h_{o} \) has 
\( \gamma \) low lying eigenvalues 
which are separated from larger eigenvalues
by a finite gap.

In what follows, we shall explicitly find
these low lying eigenvalues
(all of which will turn out to be equal to \( -s(1+m\,\nu^2) \)),
and characterize all the corresponding eigenstates.

For \( \nu>0 \), we define
\( \Gn \)
as the space of all
\( \Phi\in\Hfn \) which are written as linear combinations
of \( \Psi_{2}^{(\nu)}(\Eup,\Edn) \)
such that \( \Eup\cap\Edn\ni o \).
\( \Gn \) is not empty since
\( a^\dagger_{o,\sigma}\vac\in\Gn \).

In other words, \( \Gn \) is a set of all
\( \Phi \) which are expanded as
(\ref{eq:Phiexp2}) with the coefficients
\( \varphi(\Eup,\Edn) \) satisfying
the conditions
(\ref{eq:zecp})
and an additional condition that
\( \varphi(\Eup,\Edn)=0 \) unless
\( \Eup\cup\Edn\ni o \).
Again we see that the set of allowed
coefficients \( \{\varphi(\Eup,\Edn)\} \)
is independent of \( \nu \).
Since the basis states 
\( \Psi_{2}^{(\nu)}(\Eup,\Edn) \)
are mutually linear independent for each fixed
\( \nu\ge0 \),
we find that \( \Gn \) for different \( \nu\ge0 \)
are all identical as linear spaces.
In particular \( \Gn \)
for any \( \nu>0 \) has the same dimension 
as the space
\( \Gz \), i.e., \( \gamma \).

Note that any \( \Phi\in\Gn \)
can be written uniquely in the form
\begin{equation}
	\Phi=a^\dagger_{o,\up}\Phi_{1}
	+a^\dagger_{o,\dn}\Phi_{2},
	\label{eq:thePhi}
\end{equation}
where \( \Phi_{1} \) and
\( \Phi_{2} \) are linear combinations
of \( \Psi_{2}^{(\nu)}(\Eup,\Edn) \)
with \( \Eup\cap\Edn=\emptyset \)
and
\( \Eup\cup\Edn\not\ni o \).

We will show that,
for arbitrary 
\( t \), \( U \), and \( \nu \),
any \( \Phi\in\Gn \)
is an eigenstate of 
the local Hamiltonian \( h_{o} \)
with eigenvalue
\( -s(1+m\,\nu^2) \).
Note that this eigenvalue converges
to \( -s \) as \( \nu\to0 \),
and is \( \gamma \)-fold degenerate.
These facts imply that we have precisely located
the \( \gamma \) low lying eigenvalues of
\( h_{o} \)
for sufficiently large \( t,U \) and sufficiently 
small \( \nu \).
These low lying eigenvalues turned out to be
completely degenerate, and 
forming the lowest eigenvalue.
Since \( \Phi\in\Gn \) has all the properties
declared in the lemma, this leads us to the lemma.

Let \( \Phi \) be an arbitrary state in
\( \Gn \).
It only remains to prove that
\( h_{o}\Phi=-s(1+m\nu^2)\Phi \).
By construction we have
\( b^\dagger_{u,\sigma}b_{u,\sigma}\Phi=0 \)
for any \( u\in\Io \) and \( \sigma=\up,\dn \),
and
\( n_{r,\up}n_{r,\dn}\Phi=0 \)
for any \( r\in\Lao \).
Thus we only need to show that
\begin{equation}
	\sum_{\sigma=\up,\dn}
	a^\dagger_{o,\sigma}a_{o,\sigma}
	\Phi
	=
	(1+m\nu^2)\Phi.
	\label{eq:remains}
\end{equation}
From the expression (\ref{eq:thePhi}) and
\( \{a^\dagger_{o,\sigma},a_{o,\sigma}\}
=1+m\nu^2\),
one has
\begin{equation}
	\sum_{\sigma=\up,\dn}
	a^\dagger_{o,\sigma}a_{o,\sigma}
	\Phi
	=
	(1+m\nu^2)\Phi+\Phi',
	\label{eq:PPprime}
\end{equation}
where
\begin{equation}
	\Phi'=a^\dagger_{o,\up}
	\rbk{
	\sum_{\sigma=\up,\dn}
	a^\dagger_{o,\sigma}a_{o,\sigma}
	}\Phi_{1}
	+
	a^\dagger_{o,\dn}
	\rbk{
	\sum_{\sigma=\up,\dn}
	a^\dagger_{o,\sigma}a_{o,\sigma}
	}\Phi_{2}.
	\label{eq:thePhip}
\end{equation}

On the other hand, 
from the expression (\ref{eq:thePhi})
and the zero energy condition
(\ref{eq:zec}),
one has
\begin{eqnarray}
	c_{u,\up}c_{u,\dn}\Phi
	&=&
	c_{u,\up}c_{u,\dn}a^\dagger_{o,\up}\Phi_{1}
	+
	c_{u,\up}c_{u,\dn}a^\dagger_{o,\dn}\Phi_{2}
	\ret
	&=&
	a^\dagger_{o,\up}c_{u,\up}c_{u,\dn}\Phi_{1}
	+
	a^\dagger_{o,\dn}c_{u,\up}c_{u,\dn}\Phi_{2}
	+\nu
	(c_{u,\dn}\Phi_{1}-c_{u,\up}\Phi_{2})
	\ret
	&=&0,
	\label{eq:ccPlong}
\end{eqnarray}
where we used (\ref{eq:axs}).
By operating 
\( a^\dagger_{o,\up}a^\dagger_{o,\dn} \)
from the left, the final two lines
yield the relation
\begin{equation}
	a^\dagger_{o,\up}a^\dagger_{o,\dn}
	(c_{u,\dn}\Phi_{1}-c_{u,\up}\Phi_{2})=0,
	\label{eq:aacPcP}
\end{equation}
for any \( u\in\Io \).

By recalling that
\( a_{o,\sigma}=
c_{o,\sigma}-\nu\sum_{u\in\Io}c_{u,\sigma} \)
and noting that
\( c_{o,\sigma}\Phi_{1}=c_{o,\sigma}\Phi_{2}=0 \),
we can rewrite (\ref{eq:thePhip}) as
\begin{eqnarray}
	\Phi'
	&=&
	a^\dagger_{o,\up}a^\dagger_{o,\dn}a_{o,\dn}
	\Phi_{1}
	+
	a^\dagger_{o,\dn}a^\dagger_{o,\up}a_{o,\up}
	\Phi_{2}
	\ret
	&=&
	-\nu\,a^\dagger_{o,\up}a^\dagger_{o,\dn}
	\sum_{u\in\Io}
	(c_{u,\dn}\Phi_{1}-c_{u,\up}\Phi_{2})
	\ret
	&=&0,
	\label{eq:Phip0}
\end{eqnarray}
where we used (\ref{eq:aacPcP}).
Recalling (\ref{eq:PPprime}),
this completes the proof of the lemma.

\par\bigskip
It is a pleasure to thank 
Akinori Tanaka for pointing out crucial flaws 
in the earlier versions of the present work, 
and for indispensable discussions and comments.
I also wish to thank
Tom Kennedy,
Tohru Koma,
Kenn Kubo,
Koichi Kusakabe,
Elliott Lieb,
Andreas Mielke,
Bruno Nachtergaele,
Teppei Sekizawa,
and 
Shun-Qing Shen
for various useful conversations and discussions.

\end{document}